%% file: toc2018.tex
\pdfoutput=1

\documentclass[10pt, journal, compsoc, twoside, hidelinks]{IEEEtran}
\usepackage[utf8]{inputenc}
\usepackage[T1]{fontenc}
\usepackage[english]{babel}

\usepackage[nocompress]{cite}

\newcommand{\ignore}[1]{}
\usepackage{fancyhdr}
\usepackage[normalem]{ulem}
\usepackage[hyphens]{url}
\usepackage{booktabs}
\usepackage{xfrac}
\usepackage{algorithm}
\usepackage{algorithmic}
\usepackage{xspace}
\usepackage{tabularx}
\usepackage{placeins}
\usepackage[para]{threeparttable}
\usepackage{multirow}
\usepackage{multicol}
\usepackage{color}
\usepackage{tikz}
\usepackage[binary-units]{siunitx}
\usepackage{amssymb}
\usepackage{pifont}
\usepackage[acronym]{glossaries}
\usepackage[inline]{enumitem}
\usetikzlibrary{arrows,positioning}

\usepackage{colortbl}

\usepackage{amsmath}
\usepackage{hyperref}
\usepackage{listings}
\usepackage{etoolbox}
\usepackage[printwatermark]{xwatermark}

\definecolor{yellow}{rgb}{1.0,1.0,0.5}
\definecolor{darkorange}{rgb}{1.0,0.2,0.0}
\definecolor{cobalt}{rgb}{0.0, 0.28, 0.67}
\definecolor{darkblue}{rgb}{0.0, 0.0, 0.55}
\definecolor{lightblue}{rgb}{0.0, 0.4, 0.8}
\definecolor{darkgreen}{rgb}{0.00, 0.44, 0.0}
\definecolor{forestgreen}{rgb}{0.0, 0.27, 0.13}
\definecolor{internationalorange}{rgb}{1.0, 0.31, 0.0}
\definecolor{revfgcolor}{rgb}{0.33, 0.66, 0.0}
\definecolor{revbgcolor}{rgb}{0.90, 1.0, 0.85}

\def\revcolorname{lightblue}
\newcommand{\revcolor}{\color{\revcolorname}}

\newcommand{\revb}[1]{#1}

\newcommand{\rev}[1]{#1}
\newenvironment{revised}{}{}

\newcolumntype{C}{>{\centering\arraybackslash}X}
\newcolumntype{R}{>{\raggedleft\arraybackslash}X}
\newcolumntype{x}[1]{>{\centering\arraybackslash\hspace{0pt}}p{#1}}

\input{common}

\makeglossaries
\input{acronyms}

\begin{document}
%
\title{A Scalable Near-Memory Architecture for Training Deep Neural Networks on Large In-Memory Datasets}
%
%
%
%

\author{Fabian~Schuiki,
    Michael~Schaffner,
    Frank~K.~Gürkaynak,
    and~Luca~Benini,~\IEEEmembership{Fellow,~IEEE}
\thanks{Manuscript received 18 Feb 2018; revised 4 Aug and 26 Sep 2018.}
}

%
%

\markboth{
  IEEE Transactions on Computers,~Vol.~{(vol)}, No.~{(no)}, {(month)}~{(year)}
}{%
  \ifCLASSOPTIONpeerreview\else Schuiki \MakeLowercase{\textit{et al.}}: \fi
  A Scalable Near-Memory Architecture for Training Deep Neural Networks on Large In-Memory Datasets
}
%



\IEEEtitleabstractindextext{%
\begin{abstract}
Most investigations into near-memory hardware accelerators for deep neural networks have primarily focused on inference, while the potential of accelerating training has received relatively little attention so far. Based on an in-depth analysis of the key computational patterns in state-of-the-art gradient-based training methods, we propose an efficient near-memory acceleration engine called NTX that can be used to train state-of-the-art deep convolutional neural networks at scale. Our main contributions are: \rev{(i) a loose coupling of RISC-V cores and NTX co-processors reducing offloading overhead by $7\times$ over previously published results; (ii) an optimized IEEE\,754 compliant data path for fast high-precision convolutions and gradient propagation; (iii) evaluation of near-memory computing with NTX embedded into residual area on the Logic Base die of a Hybrid Memory Cube; and (iv) a scaling analysis to meshes of HMCs in a data center scenario. We demonstrate a $2.7\times$ energy efficiency improvement of NTX over contemporary GPUs at $4.4\times$ less silicon area, and a compute performance of \SI{1.2}{\TFLOPs} for training large state-of-the-art networks with full floating-point precision. At the data center scale, a mesh of NTX achieves above 95\% parallel and energy efficiency, while providing $2.1\times$ energy savings or $3.1\times$ performance improvement over a GPU-based system.}
\end{abstract}

}

\maketitle

\IEEEdisplaynontitleabstractindextext

%
\IEEEpeerreviewmaketitle

\input{sec_intro}
\input{sec_arch}
\input{sec_progmod}
\input{sec_results}
\input{sec_relwork}
\input{sec_conc}

\begin{revised}
\section*{Acknowledgments}

The authors would like to thank Lukas Cavigelli, Derek Chou, and Erfan Azarkhish for the inspiring discussions and insights.

This work has been supported by Microsoft Research under the project ``Enabling Practical, Efficient and Large-Scale Computation Near Data to Improve the Performance and Efficiency of Data Center and Consumer Systems'' with MRL contract number 2017-044.

This work has received funding from the European Union’s Horizon 2020 research and innovation programme under grant agreement No 732631, project ``OPRECOMP''.
\end{revised}

\bibliographystyle{IEEEtran}
\bibliography{IEEEabrv,ref}



%


\begin{IEEEbiography}[{\includegraphics[width=1in,height=1.25in,clip,keepaspectratio]{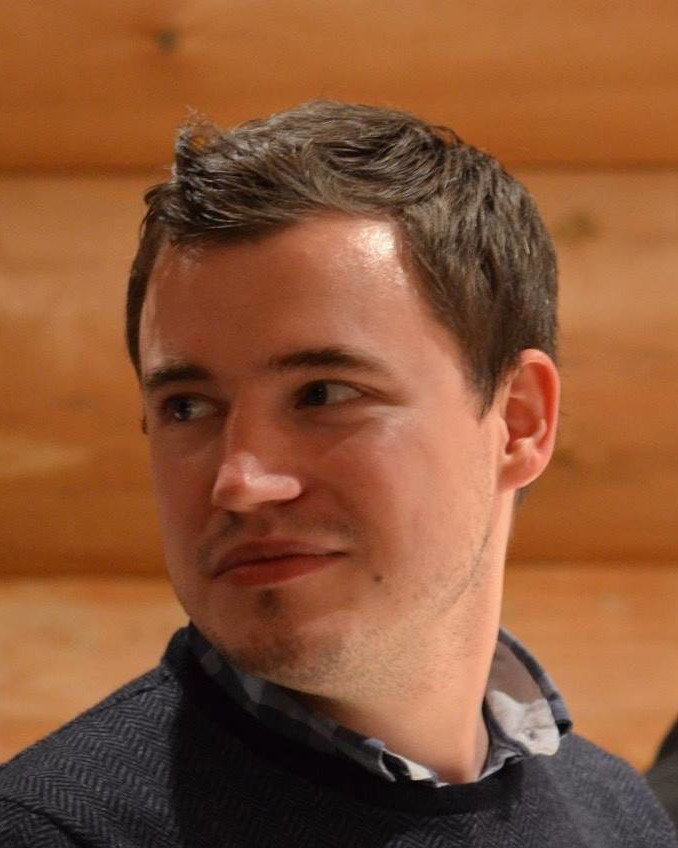}}]{Fabian Schuiki}
  received the B.Sc. and M.Sc. degree in electrical engineering from ETH Zürich, in 2014 and 2016, respectively. He is currently pursuing a Ph.D. degree with the Digital Circuits and Systems group of Luca Benini. His research interests include transprecision computing as well as near- and in-memory processing.
\end{IEEEbiography}
\vspace{-7mm}

\begin{IEEEbiography}[{\includegraphics[width=1in,height=1.25in,clip,keepaspectratio]{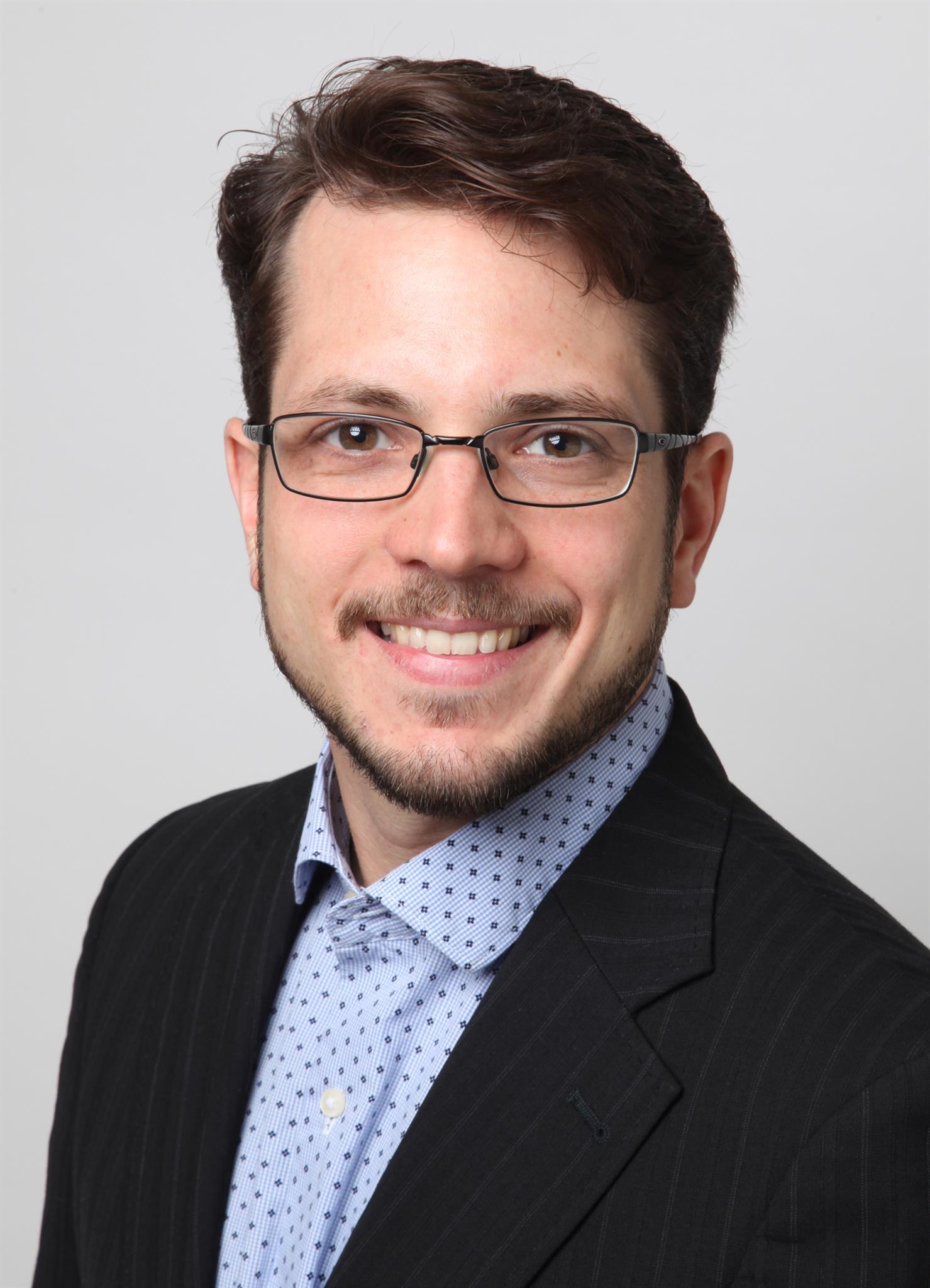}}]{Michael Schaffner}
  received his MSc and PhD degrees from ETH Zurich, Switzerland, in 2012 and 2017. He has been a research assistant at the Integrated Systems Laboratory and Disney Research from 2012 to 2017, and he is currently working as a postdoctoral researcher at the Integrated Systems Laboratory. His research interests include digital signal and video processing, and the design of VLSI circuits and systems. Michael Schaffner received the ETH Medal for his Diploma thesis in 2013.

\end{IEEEbiography}
\vspace{-7mm}

\begin{IEEEbiography}[{\includegraphics[width=1in,height=1.25in,clip,keepaspectratio]{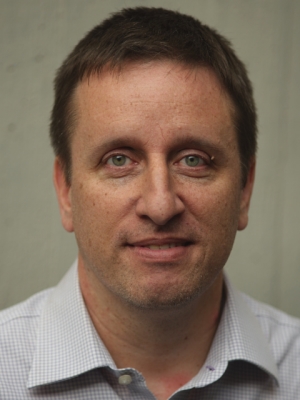}}]{Frank K. Gürkaynak}
  received the B.Sc. and M.Sc. degrees in electrical engineering from Istanbul Technical University, and the Ph.D. degree in electrical engineering from ETH Zürich, in 2006. He is currently a Senior Researcher with the Integrated Systems Laboratory at ETH Zürich, where his research interests include digital low-power design and cryptographic hardware.
\end{IEEEbiography}
\vspace{-7mm}

\begin{IEEEbiography}[{\includegraphics[width=1in,height=1.25in,clip,keepaspectratio]{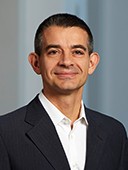}}]{Luca Benini}
  received the Ph.D. degree in electrical engineering from Stanford University, CA, USA, in 1997. He has held visiting and consulting researcher positions at EPFL, IMEC, Hewlett-Packard Laboratories, and Stanford University. His research interests are in energy efficient system design and multi-core SoC design. He is a member of the Academia Europaea, a Full Professor at the University of Bologna, and Chair of the Digital Circuits and Systems group at ETH Zürich.
\end{IEEEbiography}




\end{document}

%% file: common.tex
\newcommand{\ubold}{\fontseries{b}\selectfont}
\robustify\ubold
\newcolumntype{k}[1]{S[
  table-format=#1,
  detect-weight,
  input-symbols={()},
]}

\newcommand{\secref}[1]{{Section~\ref{#1}}}

\newcommand{\figref}[1]{{Figure~\ref{#1}}}
\newcommand{\tabref}[1]{{Table~\ref{#1}}}

\newcommand{\ifrac}[2]{{#1}\:/\:{#2}}

\DeclareSIUnit\op{op}

\DeclareSIUnit\OP{op}
\DeclareSIUnit\OPs{\OP\per\s}
\DeclareSIUnit\OPsW{\OP\per\s\per\watt}
\DeclareSIUnit\GOPs{\giga\OPs}
\DeclareSIUnit\GOPsW{\giga\OPsW}
\DeclareSIUnit\TOPs{\tera\OPs}
\DeclareSIUnit\TOPsW{\tera\OPsW}

\DeclareSIUnit\FLOP{flop}
\DeclareSIUnit\FLOPs{\FLOP\per\s}
\DeclareSIUnit\FLOPsW{\FLOP\per\s\per\watt}
\DeclareSIUnit\GFLOPs{\giga\FLOPs}
\DeclareSIUnit\GFLOPsW{\giga\FLOPsW}
\DeclareSIUnit\TFLOPs{\tera\FLOPs}
\DeclareSIUnit\TFLOPsW{\tera\FLOPsW}

\graphicspath{{figures/}}

%% file: acronyms.tex
\newacronym{agu}{AGU}{Address Generation Unit}
\newacronym{mmu}{MMU}{Memory Management Unit}
\newacronym{tlb}{TLB}{Translation Look-aside Buffer}
\newacronym{asmc}{ASMC}{Application Specific Memory Cube}
\newacronym{cnn}{CNN}{Convolutional Neural Network}
\newacronym{dag}{DAG}{Directed Acyclic Graph}
\newacronym{dnn}{DNN}{Deep Neural Network}
\newacronym{fdsoi}{FD-SOI}{Fully Depleted Silicon On Insulator}
\newacronym{flops}{flops}{Floating-Point Operations}
\newacronym{fmac}{FMAC}{Floating-Point Multiply and Accumulate}
\newacronym{fma}{FMA}{Fused Multiply and Add}
\newacronym{gru}{GRU}{Gated Recurrent Unit}
\newacronym{hbm}{HBM}{High Bandwidth Memory}
\newacronym{hmc}{HMC}{Hybrid Memory Cube}
\newacronym{lim}{LiM}{Logic in Memory}
\newacronym{lob}{LoB}{Logic Base}
\newacronym{lstm}{LSTM}{Long Short-Term Memory}
\newacronym{mlp}{MLP}{Multi-Layer Perceptron}
\newacronym{pcs}{PCS}{Partial Carry-Save}
\newacronym{pim}{PiM}{Processor-in-Memory}
\newacronym{pue}{PUE}{Power Usage Effectiveness}
\newacronym{relu}{ReLU}{Rectified Linear Unit}
\newacronym{rnn}{RNN}{Recurrent Neural Network}
\newacronym{rtl}{RTL}{Register Transfer Level}
\newacronym{sgd}{SGD}{Stochastic Gradient Descent}
\newacronym{tcdm}{TCDM}{Tightly-Coupled Data Memory}
\newacronym{mac}{MAC}{Multiply and Accumulate}
\newacronym{vfs}{VFS}{Voltage and Frequency Scaling}
\newacronym{rmse}{RMSE}{Root Mean Squared Error}
\newacronym{hwl}{HWL}{Hardware Loop}
\newacronym{fp}{FP}{floating-point}
\newacronym{fpu}{FPU}{floating-point unit}

%% file: sec_intro.tex
\IEEEraisesectionheading{\section{Introduction}\label{sec:intro}}

\begin{figure}
  \centering
  \includegraphics[width=\linewidth]{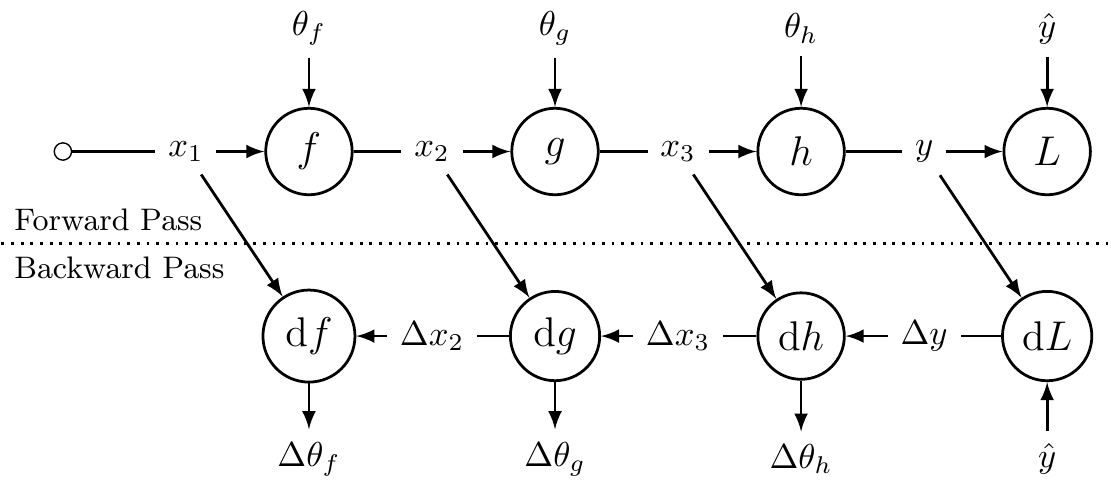}
  \caption{Data dependency graph of the forward pass (above) and backward pass (below). $f,g,h$ are \gls{dnn} layers, $L$ is the loss function, $x_1,x_2,x_3$ and $\theta_1,\theta_2,\theta_3$ are the layer activations and parameters. The backward pass introduces a data dependency between the first node $f$ and the last node $\mathrm{d} f$. Thus intermediate activations need to be stored.}
  \label{fig:intro:flow}
\end{figure}

\IEEEPARstart{M}{odern} \glspl{dnn} have to be trained on clusters of GPUs and millions of sample images to be competitive \cite{Szegedy2015}. Complex networks can take weeks to converge during which the involved compute machinery consumes megajoules of energy to perform the exa-scale amount of operations required. Inference, i.e. evaluating a network for a given input, provides many knobs for tuning and optimization. Substantial research has been performed in this direction and many good hardware accelerators have been proposed to improve inference speed and energy efficiency \cite{Sze2017}. \rev{The training of \glspl{dnn} is much harder to do and many of these optimizations do no longer apply. \Gls{sgd} is the standard algorithm used to train such deep networks \cite{Goodfellow2016}. Consider \figref{fig:intro:flow} which shows the data dependencies when training a simple neural network. While inference is concerned only with finding $y$, training aims at finding the gradients ($\Delta\theta$) which introduces a data dependency that requires us to temporarily store the output $x_i,y$ of \emph{every} layer. This also prevents optimizations such as fusing activation or sub-sampling functions with the preceding layer.}

\begin{revised}
While it has been shown that inference is robust to lowering arithmetic precision \cite{Sze2017}, the impact of fixed-point or reduced-precision \gls{fp} arithmetic on training is not yet fully understood (see \secref{sec:relwork}). Until additional light is shed on the topic, a training accelerator must support 32\,bit \gls{fp} arithmetic to be able to compete with the ubiquitous GPU. Existing accelerators require a significant amount of custom silicon and often additional memory and computational resources to function. In this paper we show that a processing system embedded in the \gls{lob} of a \gls{hmc} is a competitive and scalable option for training \glspl{dnn} in the data center. \revb{The proposed architecture is based on the earlier NeuroStream (NS) \cite{azarkhish2017neurostream} inference engine which introduced the concept of streaming coprocessors based on nested hardware loops and address generators to \gls{hmc}. We show that such corprocessors can be extended to training workloads and be made more efficient by increasing their level of autonomy. High overall data center-level energy efficiency can be achieved by distributing training over multiple such \glspl{hmc}.} The key contributions of this paper are:

\begin{enumerate}
	\item A compute architecture featuring a few RISC-V cores loosely coupled with several NTX co-processors (1:8 ratio) capable of managing computation and L1 memory access. One RISC-V core can manage 8 NTX with a reduced number of instructions, hence the von Neumann bottleneck is relaxed without compromising flexibility (\secref{sec:arch}).

	\item An optimized data path in the NTX for high-precision convolutions and gradient propagation coupled with an effective TCDM/DMA data transfer hardware that eliminates area and power overhead of large caches, by leveraging the predictability of \gls{dnn} memory patterns (\secref{sec:progmod}).

	\item Significant computational capabilities at no additional silicon area cost in the \gls{lob} of a \gls{hmc}, which we show to outperform GPUs and other accelerators in terms of silicon and energy efficiency (\secref{sec:results}).

	\item A competitive scaling to meshes of \glspl{hmc} that can replace existing GPU-based solutions in a data center setting, the improved efficiency of which translates to an increase in computational power and significant savings in power, cooling, and equipment cost (\secref{sec:results}).
\end{enumerate}
\end{revised}

The remainder of this paper is organized as follows: \rev{\secref{sec:arch} describes the proposed hardware architecture and \secref{sec:progmod} shows the execution model of \gls{dnn} layers.} \secref{sec:results} presents experimental results and comparisons to other accelerators. The remaining sections describe related and future work, and provide a conclusion.

%% file: sec_arch.tex
\section{Architecture}
\label{sec:arch}

\begin{figure*}
  \centering
  \includegraphics[width=0.94\linewidth]{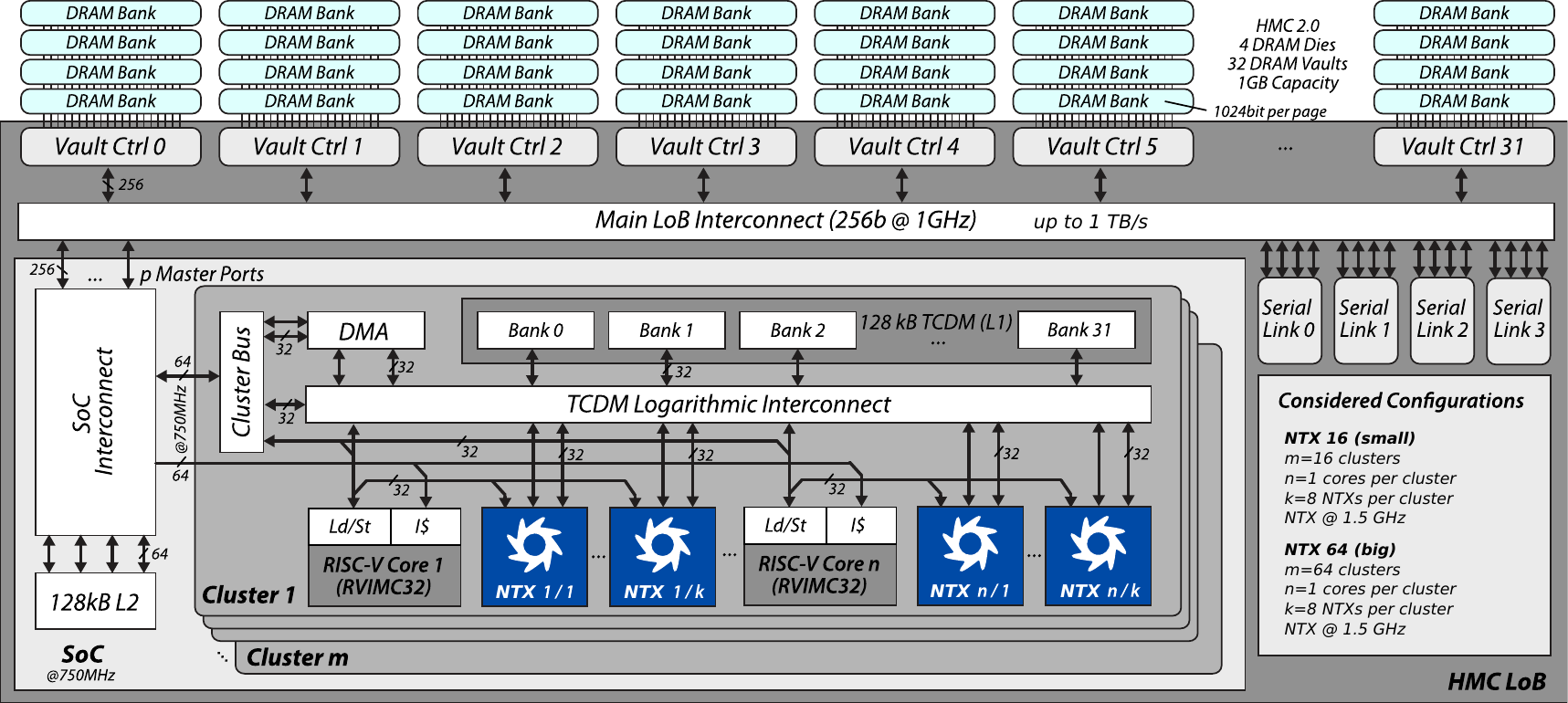}
  \caption{Top-level block diagram of one \gls{hmc} enhanced with $m$ processing clusters. The \gls{lob} contains the vault controllers, main interconnect, and the four serial links that lead off-cube. The proposed processing clusters attach directly to the main interconnect and gain full access to the \gls{hmc}'s memory space and the serial links. Each cluster consists of a DMA unit, a \gls{tcdm}, and one or more RISC-V processor cores augmented with NTX streaming co-processors. We designed the NTX to operate at \SI{1.5}{\GHz}, while the remaining additions to the system operate at \SI{750}{\MHz}.}
  \label{fig:arch:system}
\end{figure*}

\begin{figure*}
  \centering
  \includegraphics[width=0.94\linewidth]{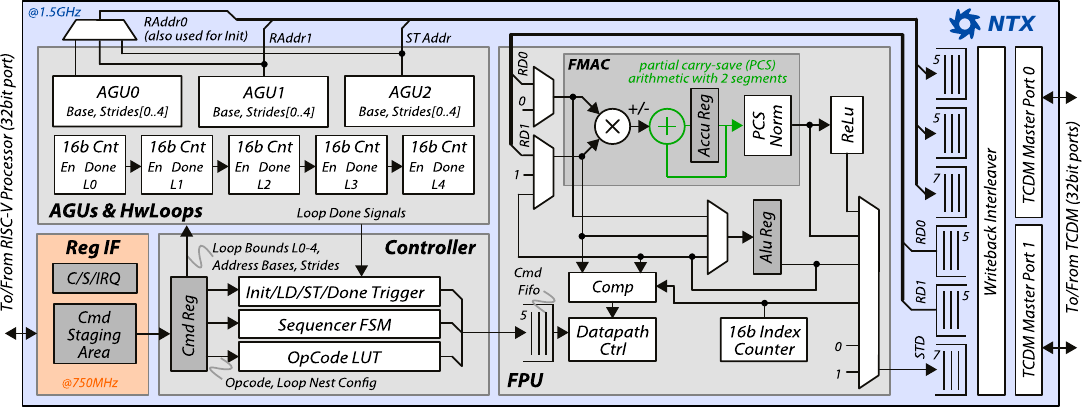}
  \caption{Block diagram of the NTX accelerator. It contains 5 hardware loops; 3 address generator units; a double-buffered command staging area in the register interface; a main controller; and a FPU with a comparator, index counter (for argmax calculations) and a fast \gls{fmac} unit. The employed depths for all FIFOs are indicated and have been determined in simulations for a TCDM read-latency of 1 cycle.}
  \label{fig:arch:nsl}
\end{figure*}

The \gls{lob} of a \gls{hmc} offers a unique opportunity to introduce a \gls{pim} as depicted in \figref{fig:arch:system}. The memory dies are subdivided into vertically connected vaults, with individual memory controllers on the \gls{lob}. Traffic between the serial links and vaults is transported by the means of an \rev{efficient all-to-all network \cite{Azarkhish2016b, HmcV21}}. Our architecture consists of multiple processing clusters attached to a crossbar, which thus gain full access to the entire memory space of the \gls{hmc}. \revb{The architecture was chosen to offer the same bandwidth as in \cite{azarkhish2017neurostream} such that the interconnect offers the full bandwidth required by the aggregate cluster ports.} The memory cube is attached to a host CPU or other cubes via the four serial links. The on-chip network is responsible for arbitration of traffic between the serial links, the DRAM, and the \gls{pim}. This arbitration can be prioritized such that external memory accesses from the serial links are given priority over internal ones originating in the processing system. It also allows requests from the \gls{pim} to be routed to the serial links for inter-HMC communication.

\subsection{Processing Cluster}
\label{sec:arch:cluster}

We combine a general purpose RISC-V processor core \cite{Gautschi2017} with multiple NTX \gls{fp} streaming co-processors. Both operate on a \SI{128}{\kilo\byte} \gls{tcdm} which offers a shared memory space with single-cycle access. The memory is divided into 32 banks that are connected to the processors via a low-latency logarithmic interconnect. These form a cluster which also contains a DMA engine that is capable of transferring two-dimensional planes of data between the \gls{tcdm} and the \gls{hmc}'s memory space. This solution has proven to be more area and energy efficient than implicit caches, and the DMA can anticipate and time block data transfers precisely, thereby hiding latency. The RISC-V processors perform address calculation and control data movement via the DMA. Actual computation is performed on the data in the \gls{tcdm} by the NTX co-processors which we describe in the next section.

Address translation is performed either in software or via a lean \gls{mmu} with \gls{tlb} as described in \cite{Azarkhish2016b}. This allows the \gls{pim} to directly operate on virtual addresses issued by the host. If there are multiple \glspl{hmc} attached to the host, care must be taken since the \glspl{pim} can only access the memory in the \gls{hmc} that they reside in. An additional explicitly managed memory outside the clusters, labeled ``L2'' in \figref{fig:arch:system}, holds the RISC-V binary executed by the processors and additional shared variables. The binary is loaded from DRAM.

\subsection{Network Training Accelerator (NTX)}
\label{sec:arch:nsl}

\begin{figure*}
  \centering
  \includegraphics[width=0.94\linewidth]{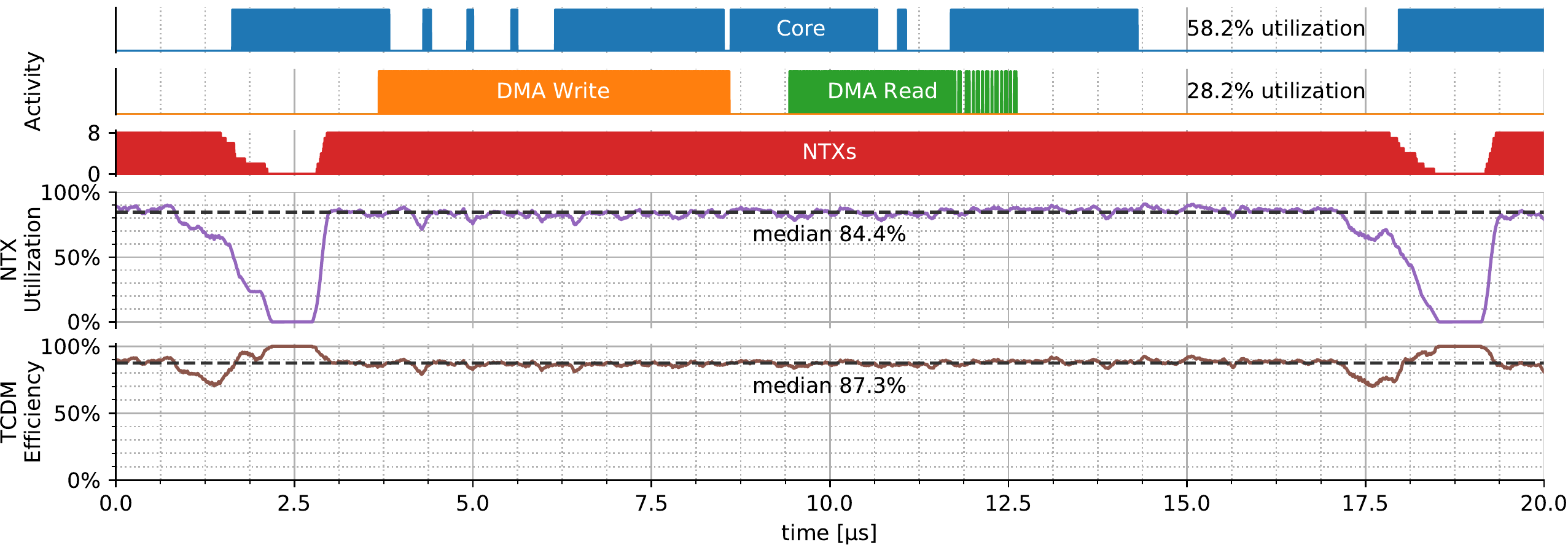}
  \caption{A $3\times3$ convolution running on one cluster. The periods of activity of the RISC-V processor and the DMA unit are shown as blocks, the activity of the co-processors is indicated as the number of active NTXs. The processor and DMA are busy during 58.2\% and 28.2\% of the computation, respectively. The utilization of the co-processors is given as percent of maximum throughput. The efficiency of the \gls{tcdm} is given as the percentage of memory requests serviced per cycle; the remaining requests stall due to conflicts. The system has a banking factor of 1.8.}
  \label{fig:arch:conv_cycles}
\end{figure*}

\begin{figure}
  \centering
  \includegraphics[width=0.94\linewidth]{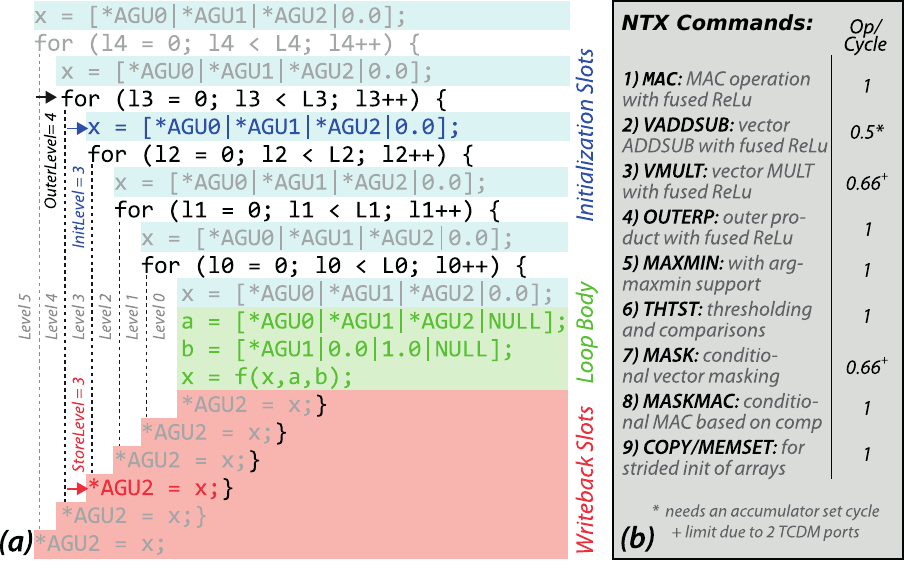}
  \caption{The structure of nested loops in C code that can be directly offloaded to NTX (a), and an overview of the supported commands and their throughput (b).}
  \label{fig:arch:nsl:hwl}
\end{figure}


The computations involved in training \glspl{dnn} are highly regular. To leverage this feature we developed NTX, a \gls{fp} streaming co-processor that operates directly on the \gls{tcdm}. Conceptually the NTX co-processor is similar to the one presented in \cite{azarkhish2017neurostream}, but it is a complete redesign optimized for performance and training. The streaming nature of the co-processor alleviates the need for a register file \rev{and corresponding load/store instructions.} The architecture of NTX is depicted in \figref{fig:arch:nsl}. It consists of four main blocks: (i) the FPU containing the main data path, (ii) the register interface for command offloading, (iii) the controller that decodes the commands and issues micro-instructions to the FPU, and (iv) the address generators and hardware loops.

\subsection{FMAC and FPU}
\label{sec:arch:nsl:fmac}

\input{tables/fmac_precision}

The FPU in NTX can perform fast \gls{fmac} operations with single-cycle throughput. It is based on a \gls{pcs} accumulator which aggregates the \SI{48}{\bit} multiplication result at full fixed-point precision ($\SI{\approx 300}{\bit}$). After accumulation the partial sums are reduced in multiple pipelined segments. In order to reach an operating frequency above \SI{1.5}{\GHz} in \SI{28}{\nano\metre} (SS 125$^\circ$C 1.0\,V), two segments are sufficient. The employed format has been aligned with IEEE\,754 32\,bit floats. \rev{The wide accumulator and deferred rounding allows NTX to achieve higher precision than conventional \glspl{fpu} for reduction operations such as convolutions. Analysis has shown that in a full $3\times3$ convolution layer of GoogLeNet \cite{Szegedy2015} the \gls{rmse} of NTX is $1.7\times$ lower than that of a 32\,bit \gls{fpu}, with respect to a common baseline (64\,bit float). See \tabref{tbl:fmac_precision}.}

\rev{The \gls{fmac} unit allows NTX to compute inner/outer product and vector addition/multiplication. An additional comparator, index counter, and ALU register enable various additional commands such as finding minima/maxima, ReLU, thresholding and masking, and copy/memset.}

\subsection{Hardware Loops and Address Generation}
\label{sec:arch:nsl:hwl}

\begin{revised}
At the core of address generation in NTX are the five \glspl{hwl}. Each loop is managed by a 16\,bit counter that has a programmable maximum count register $N_i$. Additionally, the counter can be explicitly enabled or disabled, and it has a signal indicating whether the counter has reached its maximum value and is about to reset to zero. To support nesting, each counter is enabled by the previous counter's "done" signal. The first counter (L0) is only disabled upon a pipeline stall. The "done" signal of the last counter (L4) indicates that all loop iterations have been performed. The enable signals of all counters are concatenated into a 5\,bit output.

Three \glspl{agu} allow NTX to keep track of three pointers into memory. Each unit consists of a 32\,bit register holding the address and an adder. The address is incremented by one of five programmable step sizes $p_i$, each of which corresponds to one of the hardware loops. The enabled counter with the highest index dictates the chosen stride. This allows addresses of the form
\begin{align}
A &= A_\text{base} + i_0 s_0 + i_1 s_1 + i_2 s_2 + i_3 s_3 + i_4 s_4
\end{align}
to be calculated, but using only one addition per cycle. The conversion from strides $s_i$ to step sizes $p_i$ is trivial:
\begin{align}
p_0 &= s_0 \\
p_i &= s_i - (N_{i-1}-1) \cdot p_{i-1}
\end{align}
where $N_i$ is the iteration limit of an \gls{hwl}. This conversion can be performed by the controlling CPU core when programming a command, for example as part of a driver library.

\figref{fig:arch:nsl:hwl}a shows the pseudo code structure of nested loops that NTX can natively perform. The number of loops (\emph{outer level}), position of the accumulator initialization (\emph{init level}), and position of the accumulator write back (\emph{store level}) are fully programmable. The \glspl{agu} provide addresses for the memory reads and writes depicted. The operation performed by the FPU always occurs in the innermost loop body and can be configured to be one of the commands listed in \figref{fig:arch:nsl:hwl}b.
\end{revised}

\input{tables/offloading_costs}

\begin{revised}
\subsection{Offloading Support}
\label{sec:arch:nsl:offloading}

Offloading to NTX has been enhanced with respect to the earlier NeuroStream (NS) \cite{azarkhish2017neurostream} inference engine to significantly improve efficiency in training workloads. The three improvements in offloading are: (i) a command staging area, (ii) an increased number of hardware loops, and (iii) a third address generator. The following paragraphs briefly explains the impact of each.

(i) In NS, configuration of addresses, strides, loops, and the initialization of the accumulator are performed via a command register. This register is mapped into the controlling CPU's memory space and the written commands are pushed into a FIFO. This allows the CPU to enqueue configuration updates while a computation is still ongoing. The computation itself is also a command which is popped off the FIFO only upon completion. This also implies that the FIFO needs to be deep enough to hold all commands necessary to configure the next computation, lest the CPU has to stall. NS used depth 8 for these FIFOs, causing the CPU to stall frequently.

NTX improves on this by exposing the configuration registers as a memory-mapped "staging area". As such the CPU can directly address and modify the registers. A computation is launched by writing to the command register, which is special in the sense that it causes the entire configuration to be copied to an internal "shadow" register. This allows the CPU to immediately go ahead and configure the next operation without disturbing the current one. Furthermore, parts of the configuration that do not change between commands need not be written again since the staging area is persistent. It is worthwhile noting that the size of the staging area and its shadow copy in NTX is roughly the same as the command FIFO and the corresponding registers in NS, but the former offer significantly higher ease of use. All NTX controlled by a core are also accessible via a broadcast address, which further reduces offloading time for configuring common parameters.

(ii) We observe that a convolution as it appears in DNNs has six nested loops: three that iterate over each output pixel, and three to perform the per-pixel reduction of the input dimensions (3D input and output, 4D weights). NS offers three hardware loops, which allows the 3D per-pixel reduction to be expressed in one command. To compute the first convolution of GoogLeNet \cite{Szegedy2015} which has a $7\times7\times3$ kernel and yields a $112\times112\times64$ output, 802816 offloads need to be issued by the CPU each of which ideally takes only 147 cycles. This leaves only few cycles to coordinate DMA transfers and configure the next command, thus limiting the number of NeuroStreams that can be controlled by one CPU.

NTX improves on this by increasing the number of hardware loops to five. This allows multiple output pixels to be calculated with one offload. For the aforementioned convolution, this translates to only 64 offloads that need to be issued, each of which ideally takes 1843968 cycles. In practice the size of the offloaded computation is now bounded by the tile size that fits into the TCDM, thus the CPU only needs to issue one offload per NTX per tile. This reduces the control overhead of NTX to almost zero. The CPU is now free to do more elaborate data transfers, for example issuing multiple small DMA transfers to copy slices of a tensor and performing zero-padding, thus not requiring that the data be laid out in memory in a zig-zag tiling fashion as described in \cite{azarkhish2017neurostream}. This is an important improvement, since such a tiling cannot be maintained during training without significant data reshuffling between layers, which would severely reduce the energy efficiency and inflate bandwidth. \tabref{tbl:offloading_costs} shows this effect for select convolutions in GoogLeNet \cite{Szegedy2015}.

(iii) To allow NTX to calculate multiple pixels in the output image with one offload, we added a third address generator to maintain a pointer for autonomously writing back multiple results to memory. This in contrast to NS \cite{azarkhish2017neurostream} which requires an explicit command from the CPU to store the accumulated value.
\end{revised}

%% file: tables/fmac_precision.tex
\begin{table}
\begin{threeparttable}
\caption{\rev{Arithmetic error comparison between a conventional float32 FPU and the NTX 32\,bit FMAC unit, for a full $3\times3$ convolution layer of GoogLeNet \cite{Szegedy2015}, with respect to a common baseline (64\,bit float).}}
\label{tbl:fmac_precision}
\begin{tabularx}{\linewidth}{@{}Xlll@{}}
\toprule
    \textbf{Implementation} &
    \textbf{RMSE} &
    \multicolumn{2}{c@{}}{\textbf{Relative Error}} \\
    \cmidrule(l){3-4}
    && Maximum & Median \\
\midrule
    Intel CPU (float32) & $1.83\cdot10^{-7}$ & $5.42\cdot10^{-3}$ & $9.40\cdot10^{-8}$ \\
    NTX (32\,bit FMAC)  & $1.08\cdot10^{-7}$ & $1.19\cdot10^{-7}$ & $5.97\cdot10^{-8}$ \\
\bottomrule
\end{tabularx}
\end{threeparttable}
\vspace{-3mm}
\end{table}

%% file: tables/offloading_costs.tex
\begin{table}
\begin{threeparttable}
\caption{\rev{Comparison of the number of offloads necessary and execution time of an offloaded command for NTX and NeuroStream (NS) \cite{azarkhish2017neurostream}, for different convolution layers of GoogLeNet \cite{Szegedy2015}. NS requires one offload per output pixel, whereas the increased number of hardware loops and the third address generation unit of NTX allow it to compute many output pixels per offloaded command.}}
\label{tbl:offloading_costs}
\begin{tabularx}{\linewidth}{@{}lX k{6.0} k{3.0} k{3.0} k{7.0} @{}}
\toprule
    \textbf{Kernel} &
    \textbf{Output} &
    \multicolumn{2}{c}{\textbf{Offloads}} &
    \multicolumn{2}{c@{}}{\textbf{Busy Cycles per Offload}} \\
    \cmidrule(lr){3-4}
    \cmidrule(l){5-6}
    && NS & NTX & NS & NTX \\
\midrule
    7x7x3       & 112x112x64  & 802816        & 64             & 147         & 1843968 \\
    3x3x64      & 56x56x192   & 602112        & 192            & 576         & 1806336 \\
    1x1x256     & 28x28x64    & 50176         & 64             & 256         & 200704  \\
    1x1x512     & 14x14x192   & 37632         & 192            & 512         & 100352  \\
\bottomrule
\end{tabularx}
\end{threeparttable}
\end{table}

%% file: sec_progmod.tex
\section{NTX Execution}
\label{sec:progmod}

\begin{revised}
The combination of RISC-V processors and dedicated \gls{fp} streaming co-processors makes our architecture very flexible. It is a many-core platform with explicitly managed scratchpad memories, where data copies are performed by a DMA engine and bulk computations by NTX co-processors in parallel to a running program. The proposed architecture allows for entire \gls{dnn} training batches to be performed completely in memory, without intervention from a host outside the \gls{hmc}, as follows. Starting from a reference implementation of a training step in C or C++, nested loops of the form described in \secref{sec:arch} are amenable to acceleration on NTX. This includes the bulk operation of all \gls{dnn} layers. These loops are replaced by an offload sequence consisting of writes to the eight staging areas. Furthermore, the input and output data of each loop nest must be tiled and data movement appropriately scheduled, as described below.

\figref{fig:arch:conv_cycles} shows the execution of one tile of a $3\times3$ convolution on NTX. The RISC-V core first configures and launches the main computation on NTX, then controls the DMA to write back the output of the previous tile and read the input of the next one. Additional tasks such as zero padding and address computation are performed in the background. The short period of NTX idleness in between tiles is due to the core using the NTX to initialize the next tile, which is a quick command that terminates faster than the core can configure the next big computation.



\end{revised}

\subsection{Memory and Tiling}
\label{sec:progmod_memory}


As described in \secref{sec:arch} and \cite{azarkhish2017neurostream, Azarkhish2016b}, \rev{the core and NTX operate directly on a scratchpad memory inside the cluster. A DMA unit in conjunction with a lean \gls{mmu} is used to copy data from DRAM into cluster memory, where the accelerators operate on them. This mechanism is similar to how caching works on CPUs/GPUs, but is explicitly managed by the programmer.}

\begin{revised}

The scratchpad memory in the cluster is limited in size. To evaluate an entire convolution layer for example, the input and output data are tiled to fit into memory. The tiles need to overlap in the convolution case. The DMA unit can run in parallel to the computation and is used to write back previous results and read next inputs while a computation is ongoing (double buffering), as can be seen in \figref{fig:arch:conv_cycles}. In \cite{azarkhish2017neurostream} the authors made use of 4D tiling, which requires that the data is already laid out in such a tiled fashion in DRAM, including replication of the overlapping areas. This allows the DMA to copy a tile in a single consecutive transfer, requiring little control by the core. For training this scheme is infeasible, since forward and backward passes require different tile sizes, and the data would need to be retiled after several subsampling layers to maintain a sufficient tile size. This retiling translates into no-op movement of data, which wastes bandwidth and energy.

The improved offloading scheme described in \secref{sec:arch:nsl:offloading} and increased independence of NTX compared to \cite{azarkhish2017neurostream} frees up significant RISC-V core resources. This allows us to now store tensors in memory as dense chunks of \gls{fp} values, without any replication or tiling pre-applied. To transfer tiles, we task the core with issuing multiple DMA transfers, each of which copies one consecutive stripe of data. Zero padding can also be performed by the core in this way. Hence we can drop the requirement of the data being laid out in memory in a pre-tiled fashion.
\end{revised}

\begin{figure}
  \includegraphics[width=\linewidth]{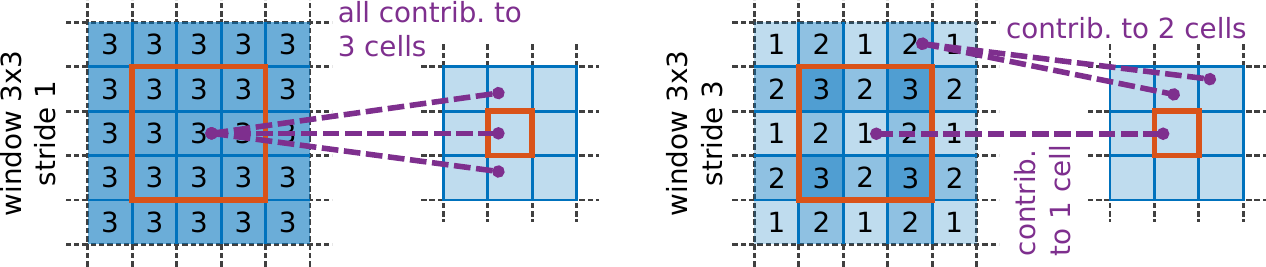}
  \caption{Irregularity introduced by stride in stencil operations such as convolution and max pooling. With a stride of one (left), all input cells contribute to the same number of output cells. With a stride of three (right), input cells contribute to one, two, or three output cells.}
  \label{fig:progmod:stride}
\end{figure}

\begin{revised}
\subsection{Strided Stencil Operations}
\label{sec:progmod_strided}

A stride greater than one in stencil operations such as convolution and pooling causes an irregularity during training. For example, a strided convolution can be thought of as a regular convolution where a subset of the output pixels are discarded. The backward pass correspondingly can be represented roughly as a sparse convolution where the discarded pixels are 0. For efficiency reasons we would like to skip multiplications with 0, effectively leveraging the sparsity of the problem. However NTX cannot change the number of summands within the course of one operation, so we must perform convolutions that have a constant number of operations required per pixel. This does not hold for strided convolutions in the backward pass, where the input derivative contains contributions from a varying number of output pixels. See figure \figref{fig:progmod:stride}. We observe that we can subdivide the pixels of the input derivative into different categories: Each pixel subset can be computed as a regular convolution with a subset of the filter weights, and the overall result can be found by interleaving the subset results. This scheme allows us to decompose a sparse convolution (as found in the derivative of strided convolutions) into multiple dense convolutions each contributing a subset of the result pixels.




\end{revised}

\subsection{Special Functions (exp, log, div, sqrt)}
\label{sec:progmod_exp}

There is no dedicated hardware to evaluate special functions such as division, exp, log, square roots, or arbitrary powers. \rev{These are needed for the softmax layer or various forms of normalization.} As the number of such operations is typically very low (in the order of a few thousands per training step), it is feasible to implement them using iterative algorithms on the NTX, calculating multiple results in parallel. We found that for tens to hundreds of inputs, pipeline latency can be hidden and the evaluation takes on the order of 30 to 100 cycles per element.


\begin{revised}
\subsection{Communication across HMCs}
\label{sec:progmod_comms}

The serial links in the \gls{hmc} are accessible to the processor cores and DMA units in each cluster. This allows a mesh of \glspl{hmc} to be programmed in a similar way as a two-tiered network of compute nodes. Within the \gls{hmc}, clusters may exchange data via the DRAM and L2. Across \glspl{hmc}, the processor cores may cooperate to perform complex systolic operations via the serial links. \secref{sec:results:multiple_hmcs} provides an example of this.
\end{revised}

%% file: sec_results.tex
\section{Experimental Results and Analysis}
\label{sec:results}

In this section we evaluate the silicon and energy efficiency of our proposed architecture and compare it against NeuroStream, the most closely related other accelerator \cite{azarkhish2017neurostream}. Furthermore we investigate the effects of voltage and frequency scaling and the impact of multiple logic dies per memory cube. We conclude by comparing different NTX configurations against existing accelerators and evaluate the data center scale impact of our architecture.

\subsection{Methodology}
\label{sec:results:meth}

\subsubsection{DRAM Power}

We model the power consumption of the vault controllers, DRAM dies, and \gls{hmc} interconnect as the following relationship:

$$P_\text{dram}(B) = \SI{7.9}{\watt} + B \cdot \SI{21.5}{\milli\watt\second\per\giga\byte},$$

where $B$ is the requested bandwidth. We call this the ``DRAM'' power. This model is based on the observation in \cite{azarkhish2017neurostream} that \SI{7.9}{\watt} are consumed in a \SI{1}{\giga\byte} cube under no traffic into DRAM (50\,nm, see \secref{sec:results:meth:techscale}). Under an average traffic of \SI{51.2}{\giga\byte\per\second} caused by their investigated workloads, this increases to the reported $\SI{9.0}{\watt}$, a bandwidth-dependent power increase of \SI{21.5}{\milli\watt\second\per\giga\byte}. These estimates are conservative and do not consider further power-saving measures, such as \gls{vfs} or power gating of \gls{hmc} components.

\subsubsection{Cluster Power}
\label{sec:results:meth:clpwr}

\begin{figure}
  \centering
  \includegraphics[width=\linewidth]{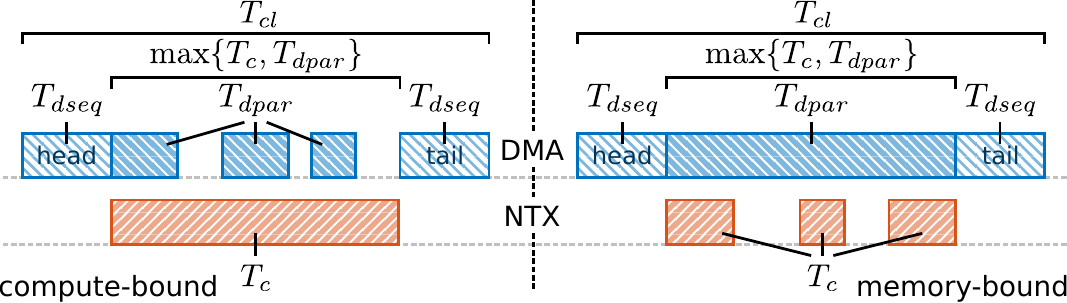}
  \caption{Execution time of a kernel running on a cluster. See \secref{sec:results:meth:clpwr} for a details. $T_\text{dseq}$ corresponds to memory transfers that need to happen before and after the main computation, e.g. first data fetch and last data store. $T_\text{dpar}$ corresponds to transfers that can happen in parallel to the computation $T_\text{c}$. Shown are a compute-bound case where $T_\text{c}$ dominates, and a memory-bound case where $T_\text{dpar}$ dominates.}
  \label{fig:results:clpwr}
\end{figure}

We have synthesized our design for a \SI{28}{\nano\meter} \gls{fdsoi} technology using Synopsys Design Compiler, which we also use to estimate power based on simulation traces. \rev{The \gls{rtl} model was back-annotated with timing information obtained from the synthesized design at \SI{125}{\celsius}/\SI{1.0}{\volt} (slow-slow corner). We execute the worst-case kernel in an RTL simulation of the cluster, which gives us a cycle-accurate picture of the computation. This is depicted in \figref{fig:arch:conv_cycles}, and furthermore gives us an estimate of the cluster's power consumption, \SI{165}{\pico\joule} per clock cycle in this case. The worst-case kernel is a convolution which makes full use of the \glspl{fpu} in the NTXs and has a high utilization of the DMA. Memory-bound kernels consume less power, since the FPU utilization is lower, reducing its power contribution. This simulation also gives us realistic utilization efficiencies of $\eta_\text{c} = 84\%$ and $\eta_\text{d} = 87\%$ for NTX and TCDM, respectively.}

\subsubsection{Network Layer Energy}
\label{sec:results:meth:layerpwr}

\rev{To evaluate applications, we model the execution of individual network layers. The computation and data movement performed by a cluster is very predictable. For each network layer we therefore compute the number of \gls{fp} operations necessary, as well as the amount of data that needs to be transferred. The latter we further split into data that must be moved before computation can start (\emph{head}), data that can be moved in parallel to the computation, and data that must be moved once the computation completes (\emph{tail}). This closely models the double buffering possible by overlapping operation of the DMA and NTX within the cluster.} For each kernel we determine the execution time of the computation ($T_\text{c}$) and DMA transfers ($T_\text{dpar},T_\text{dseq}$) as:
\begin{align}
  T_\text{c} &= \ifrac{N_\text{c}}{\eta_\text{c}\,r_\text{c}\,f} & [\si{\second}] \\
  T_\text{dpar} &= \ifrac{(D_\text{dma} - D_\text{head} - D_\text{tail})}{\eta_\text{d}\,r_\text{d}\,f} & [\si{\second}] \\
  T_\text{dseq} &= \ifrac{(D_\text{head} + D_\text{tail})}{\eta_\text{d}\,r_\text{d}\,f} & [\si{\second}]
\end{align}
where $T_\text{dpar}$ represents DMA transfers that can run in parallel with computation and $T_\text{dseq}$ those that need to happen before and after. In more detail, $N_\text{c}$ and $D_\text{dma}$ are the total number of compute operations performed and bytes transferred by the kernel; $r_\text{c}$ are the peak compute operations per cycle of the cluster; and $r_\text{d}$ is the peak bandwidth of the DMA per cycle. For the architecture with 8 NTXs presented in \secref{sec:arch}, $r_\text{c} = \SI{8}{\OP}$ and $r_\text{d} = \SI{4}{\byte}$. $\eta_\text{c}$ and $\eta_\text{d}$ account for inefficiencies such as interconnect contentions and are determined empirically from simulations. We then formulate the execution time, requested bandwidth, and power consumption of the kernel as:
\begin{align}
  T_\text{cl} &= \max\{ T_\text{c}, T_\text{dpar} \} + T_\text{dseq} & [\si{\second}] \\
  B_\text{cl} &= \ifrac{D_\text{dma}}{T_\text{cl}} & [\si{\byte\per\second}]\\
  P_\text{cl} &= \SI{165}{\pico\joule} \cdot f & [\si{\watt}]
\end{align}
See \figref{fig:results:clpwr} for a visual explanation. Note that we issue DMA transfers in chunks of multiple \si{\kilo\byte} and the engine is capable of having multiple simultaneous transfers in flight. This allows us to hide the latency into DRAM which we estimate to be on the order of 40 core cycles. It is crucial that we fully saturate the precious bandwidth into DRAM when performing strided memory accesses, e.g. when transferring a tile of a tensor. There the length of the tile's innermost dimension is critical, as it determines the length of one burst accesses. Since we have full control over the tiling, we can ensure that a tile has at least 8 elements along its shortest dimension. This yields consecutive accesses of at least \SI{32}{\byte}, which is the minimum block size in an \gls{hmc} \cite{HmcV21}.

\subsubsection{Cube Power}
\label{sec:results:meth:cubepwr}

Based on the above, we model the requested bandwidth, power consumption, and energy efficiency of a kernel parallelized on a \gls{hmc} with $K$ clusters as
\begin{align}
	B &= K \cdot B_\text{cl} & [\si{\byte\per\second}] \\
	T &= \ifrac{T_\text{cl}}{K} & [\si{\second}] \\
	P &= P_\text{dram}(B) + K \cdot P_\text{cl} & [\si{\watt}] \\
	\eta &= \ifrac{2\,N_\text{c}}{P\,T} & [\si{\FLOPsW}]
\end{align}
\rev{The parallelization is achieved by distributing the tiles of computation described in \secref{sec:progmod} across the clusters of a cube.}

\begin{revised}
\subsubsection{Network Training Energy}
\label{sec:results:meth:trainpwr}

We then model different layers of \glspl{dnn} as the amount of computation and data transfers necessary. This also gives us a per-layer estimate of the number of parameters and intermediate activations. We proceed to model each of the investigated networks as a sequence of these layers, giving us a realistic estimate for the execution time of the inference and training steps of one image on the proposed architecture, together with the associated bandwidth requirement. We then further use the execution time and the cluster and bandwidth-dependent DRAM/LoB power determined above to estimate the overall energy required to process one image.


\end{revised}

\subsubsection{Technology Scaling}
\label{sec:results:meth:techscale}

We use internal comparisons and publicly available information to estimate the effect of scaling down the technology node of the \gls{lob} from the \SI{28}{\nano\meter} \gls{fdsoi} process investigated by us to a more modern \SI{14}{\nano\meter} FinFET node \cite{tewell2016fdsoi, davis2013tsmc28vs16}. For this change we observed across several designs an increase of $1.4\times$ in speed, a decrease of $0.4\times$ in area, and $0.7\times$ in dynamic power dissipation.

To our knowledge there is no publicly available information on the DRAM characteristics of \glspl{hmc}. ``SMCSim'' \cite{Azarkhish2016b} assumes them to be similar to the MT41J512M8 device by Micron, which is based on a \SI{50}{\nano\meter} process. Given the manufacturer and \cite{Pawlowski2011}, the device seems to be a reasonable reference for early \glspl{hmc}. We estimate the DRAM technology scaling factor for power consumption to be $0.87$, by comparing the supply currents and voltages of this device to the newer \SI{30}{\nano\meter} MT40A512M8.

\begin{revised}
\subsubsection{GPU Efficiency Estimation}
\label{sec:results:meth:gpupwr}

We estimate GPU efficiency based on the training time per image measured by \cite{TfBench2017, murphy2017}. For each network we compute the amount of \si{\FLOP} necessary per image based on our model of the network. This yields an estimate of the actual throughput in \si{\FLOPs} achieved. Assuming the GPU can reach its TDP under such highly optimized workloads (e.g. cuDNN), we determine the energy efficiency as the ratio between that throughput and the TDP. We do not assume optimizations such as Winograd to be performed on the GPU, and as such overestimate the number of \gls{fp} operations performed, making the estimated energy efficiency optimistic. Furthermore, this excludes the power consumed by the CPU to constantly push training data into GPU memory.
\end{revised}

\subsection{Precision, Sparsity, Compression}

Training a \gls{dnn} with reduced \gls{fp} precision or even fixed-point arithmetic is much harder than doing the same for inference. The intuition here is that the \gls{sgd} algorithm performs smaller and smaller changes to the parameters as training progresses. If these changes fall beneath the numeric precision, the algorithm effectively stops converging. There is no a priori obvious range of magnitudes within which parameters fall, thus the arithmetic must support a significant dynamic range without additional prior analysis. NTX employs \SI{32}{\bit} \gls{fp} arithmetic which is commonly used in deep learning frameworks and CPUs/GPUs, rendering such analysis unnecessary. Note that there is evidence that training is possible in fixed-point arithmetic with little accuracy loss in some cases \cite{luo2017dadiannao}. However, results tend to be limited to specific networks and other work suggests that reducing precision may not be feasible at all without incurring significant accuracy loss \cite{koster2017flexpoint}.

Recent work on network compression and pruning techniques has shown promising results in terms of reducing computational overhead \rev{\cite{wen2016learning}}. The general purpose nature of the RISC-V processors in our architecture allows some of these schemes to be implemented. For example entire convolutions may be skipped or certain forms of decompression and re-compression may be performed on the processor cores. The NTX has not been optimized for sparse tensor operations however, and we leave their detailed analysis for future work.


\subsection{\glsfirst{vfs}}
\label{sec:results:dvfs}

\begin{figure}
  \centering
  \includegraphics[width=0.94\linewidth]{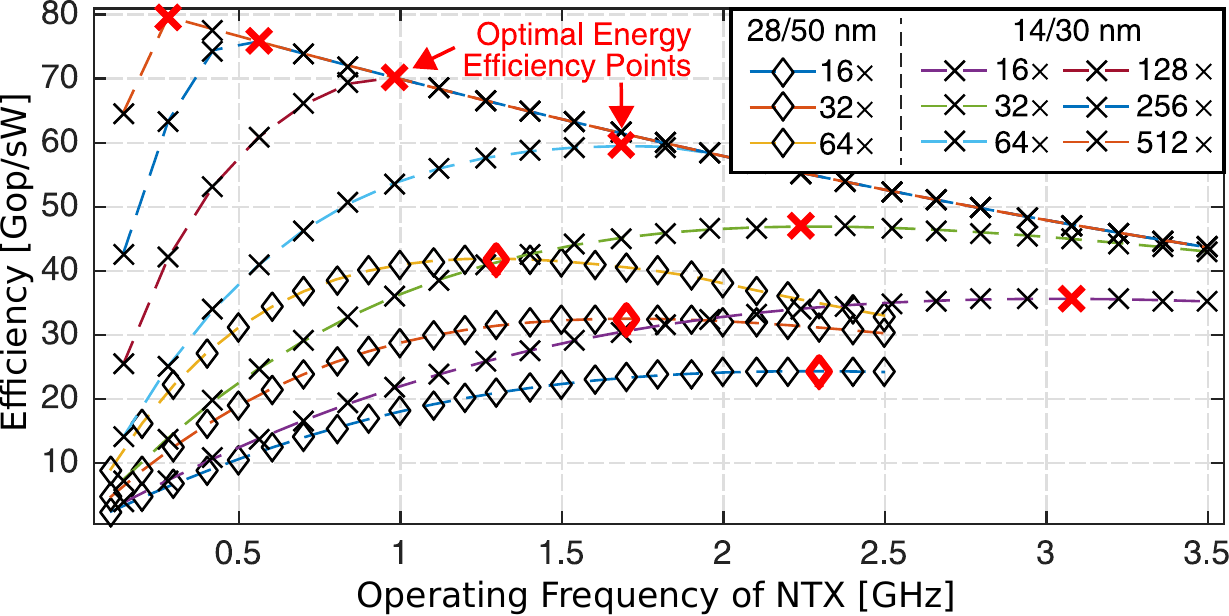}
  \caption{Energy efficiency versus operating frequency of different numbers of clusters and different technology nodes: \SI{28}{\nano\meter} logic / \SI{50}{\nano\meter} DRAM, \SI{14}{\nano\meter} logic / \SI{30}{\nano\meter} DRAM. The clusters perform a $3\times 3$ convolution. The voltage is varied between \SI{0.6}{\volt} and \SI{1.2}{\volt} in proportion to the frequency. Points of highest efficiency of each configuration are marked in bold red. The internal bandwidth of the \gls{hmc} puts an upper bound on the achievable energy efficiency, visible in the upper half of the graph.}
  \label{fig:results:eff_vs_freq}
\end{figure}

\begin{figure}
  \centering
  \includegraphics[width=\linewidth]{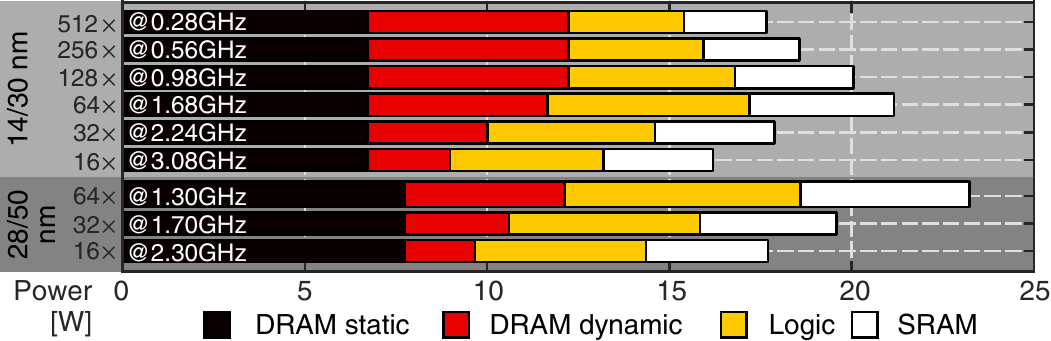}
  \caption{Power dissipation of different configurations, evaluated at their most-efficient operating point in \figref{fig:results:eff_vs_freq}. Note that even the massively parallel configurations with more than 64 clusters are below a TDP of \SI{25}{\watt}.}
  \label{fig:results:power}
\end{figure}

\begin{figure}
  \centering
  \includegraphics[width=\linewidth]{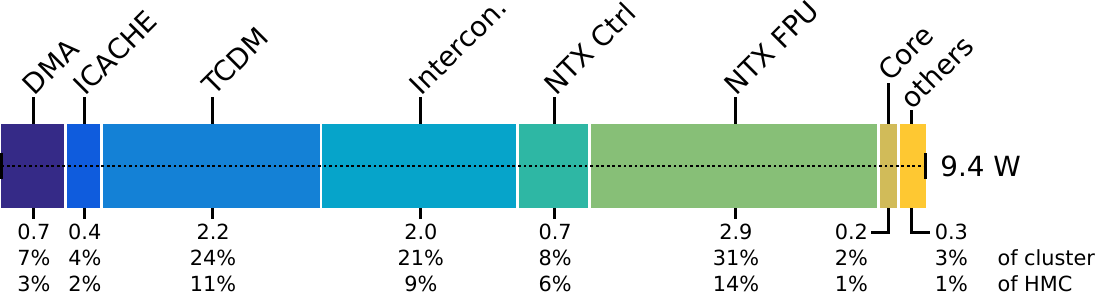}
  \caption{\rev{Power breakdown of the processing clusters in an NTX\,64 in 14\,nm performing the $3\times3$ convolution described in \figref{fig:arch:conv_cycles}. 47\% of the \gls{hmc}'s power are consumed by the clusters. More precisely, 76\% of the cluster power are dedicated to computation (NTX FPUs, TCDM, and TCDM interconnect) while 21\% are consumed by control logic (DMA, ICACHE, NTX Controller, RISC-V Core). 3\% are consumed by cluster peripherals. The DRAM, memory controllers, and interconnect in the \gls{hmc} consume another \SI{11.6}{\watt}, 53\% of the total cube power. Notably 14\% of the total \gls{hmc} power are spent in the FPUs.}}
  \label{fig:results:power_breakdown}
\end{figure}

In this section we assess the efficiency of NTX at different operating points. We vary the supply voltage between \SI{0.6}{\volt} and \SI{1.2}{\volt}; and the operating frequency between \SI{0.1}{\GHz} and \SI{2.5}{\GHz} for the \SI{28}{\nano\meter} process and \SI{0.14}{\GHz} and \SI{3.5}{\GHz} for the \SI{14}{\nano\meter} process. The voltage is assumed to scale linearly with frequency \cite{simunic2001dynamic} and is thus varied in proportion to the frequency. \figref{fig:results:eff_vs_freq} plots the energy efficiency of \glspl{hmc} with different NTX configurations against the operating frequency. Two counteracting effects lead to a tradeoff between efficiency and frequency: On one hand DRAM consumes significant static power, making it beneficial to operate at a higher frequency to decrease the time to solution. On the other hand the NTX power consumption increases quadratically with voltage and thus frequency. For larger configurations, the internal bandwidth limit of the \gls{hmc} is reached at a certain frequency, visible as a dent in the efficiency. The points of highest efficiency are listed in \tabref{tbl:results:syscmp}. \figref{fig:results:power} shows a breakdown of the power consumption at these operating points. \rev{\figref{fig:results:power_breakdown} provides a more detailed power breakdown of the NTX\,64 configuration in 14\,nm.} All configurations remain within a power budget of \SI{25}{\watt}, which according to \cite{eckert2014thermal} is feasible for a \gls{hmc} with active cooling \rev{and keeps DRAM temperature within nominal refresh limits}. If the static power of the DRAM decreases, e.g. by switching to a different memory technology, these optimal operating points will change.

\subsection{Multiple Logic Layers}
\label{sec:results:lim}

\tabref{tbl:results:syscmp} shows the area occupied by different NTX configurations. The unoccupied area on the \gls{lob} is not precisely known, and estimates range from \SI{10}{\square\milli\metre} \cite{azarkhish2017neurostream} to \SI{50}{\square\milli\metre} \cite{Gao2017}. In the following we assume that the \gls{lob} has an area of \SI{50}{\square\milli\metre}, of which \SI{25}{\square\milli\metre} are unused and thus available to custom logic. This allows configurations of up to 64 clusters per \gls{hmc}. For larger configurations, we propose the use of multiple stacked logic dies such as the 3D \gls{lim} proposed in \cite{Rush2016}. While the use of additional layers increases the complexity of the die stack, they allow for a significant increase in parallelism and efficiency. Furthermore, the use of \gls{lim} layers for custom accelerator logic has the additional benefit of decoupling the \gls{lob} manufacturing process from the accelerator, thus allowing modular assembly of ``\glspl{asmc}''. We expect this concept to be relevant for \gls{hbm} as well. 

\begin{revised}
\subsection{Memory}
\label{sec:results:mem}

\begin{figure}
  \includegraphics[width=0.94\linewidth]{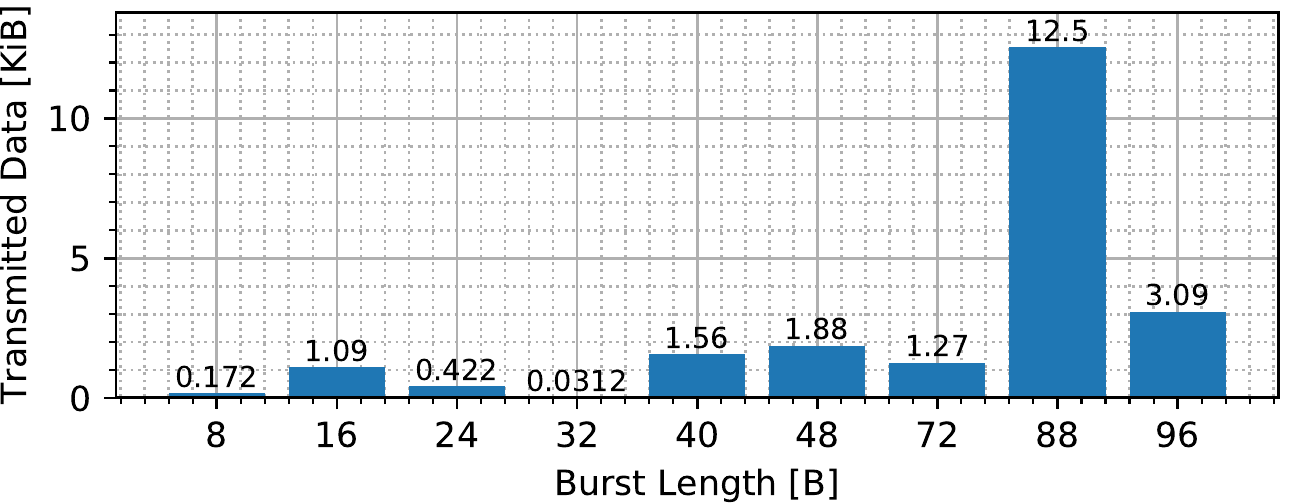}
  \caption{Histogram of data burst lengths issued by the DMA when calculating a $3\times3$ convolution tile.}
  \label{fig:arch:conv_bursts}
\end{figure}

\input{tables/network_memory}

\tabref{tbl:network_memory} summarizes our estimates for the memory occupied by the parameters and the intermediate activations of the networks investigated in the paper. We derive these from the network structure outlined in the corresponding papers. For training with a batch size of 1 the footprint amounts to \SI{239}{\mega\byte}, \SI{73.2}{\mega\byte}, and \SI{461}{\mega\byte} for AlexNet, GoogLeNet, and ResNet-152, respectively. For batch sizes greater than 1, where the gradient of each image is computed separately and added to a weighted average, another set of of parameters is needed to hold the accumulated gradient. This amounts to \SI{471}{\mega\byte}, \SI{99.8}{\mega\byte}, and \SI{767}{\mega\byte}, respectively. \revb{Note that the memory footprint then remains constant for all batch sizes.\footnote{Images in a batch may still be processed in sequence before a weight update is performed, keeping memory need constant.}} This leaves \SIrange{0.5}{7}{\giga\byte} for training data depending on network and \gls{hmc} size, around 3550 to 48600 sample images ($227\times277\times3$), equating to \SIrange{31}{247}{\second} of independent training operation on NTX~64.

To fully utilize the bandwidth into DRAM, it is pa\-ra\-mount that the accesses emitted by the DMA occur in sufficiently long bursts and have high locality with respect to DRAM pages to reduce overhead. In the case of 4D tiling \cite{azarkhish2017neurostream}, this is given by the fact that the pre-tiled data lies in DRAM as a dense consecutive sequence. In the case of on-the-fly tiling the DMA has to issue more and smaller bursts since the required data does not lie in DRAM consecutively. The tile dimensions however offer multiple degrees of freedom to adjust the access patterns generated by the clusters. For example, \glspl{hmc} \cite{HmcV21} use an internal bus width of \SI{32}{\byte}, and a maximum DRAM page size in the range of \SIrange{32}{256}{\byte}. In the aforementioned $3\times3$ convolution most data transfers occur as bursts of \SI{72}{\byte}, \SI{88}{\byte}, or \SI{96}{\byte}, and 92\% of all data is transferred in bursts above \SI{32}{\byte}. \figref{fig:arch:conv_bursts} shows a histogram of the burst lengths issued by the DMA into the DRAM. The few small bursts are due to convolution weight transfers, which can be cached to improve burst length further. We thus conclude that our architecture is capable of fully utilizing DRAM bandwidth by emitting sufficiently large accesses.
\end{revised}

\subsection{Comparison with NeuroStream}
\label{sec:results:comp_azarkhish}

\input{tables/nst_comp}

NeuroStream (NS) \cite{azarkhish2017neurostream} was aimed primarily at efficient inference and requires data to be very carefully laid out in memory (4D tiling). This constraint on data layout makes training very inefficient, since intermediate activations after each layer need to be re-tiled when storing them back to DRAM. This puts a significant workload on the RISC-V processor cores and causes additional traffic into memory. The processors are under high load to keep the NS saturated with \gls{fp} operations, such that spending compute cycles on re-tiling also means stalling the NS co-processors. Our architecture does not depend on such a tiling.

In \tabref{tbl:nst_comp} we compare NTX to NS, both implemented in 28\,nm. The much improved offloading scheme allows us to increase the ratio of co-processors to control cores from 2:1 to 8:1. The fast \gls{fmac} allows us to operate the NTX at twice the frequency of the rest of the cluster, leading to an increase of peak performance from \SI{256}{\GOPs} to \SI{384}{\GOPs} for the 16 cluster version. The increased number of hardware loops and operations supported by NTX, together with the improved performance, allow us to increase the energy efficiency of a training step from \SI{15}{\GOPsW} to \SI{21}{\GOPsW}.

The 16 cluster configuration requests a peak bandwidth of \SI{57.6}{\giga\byte\per\s}, which does not saturate the internal bandwidth of up to \SI{320}{\giga\byte} available inside the \gls{hmc}. We can improve the energy efficiency to \SI{38.3}{\GOPsW} by increasing the number of clusters to 64.

\subsection{Comparison with other Accelerators}
\label{sec:results:comp_others}

\input{tables/system_comp}

\begin{figure}
  \centering
  \includegraphics[width=\linewidth]{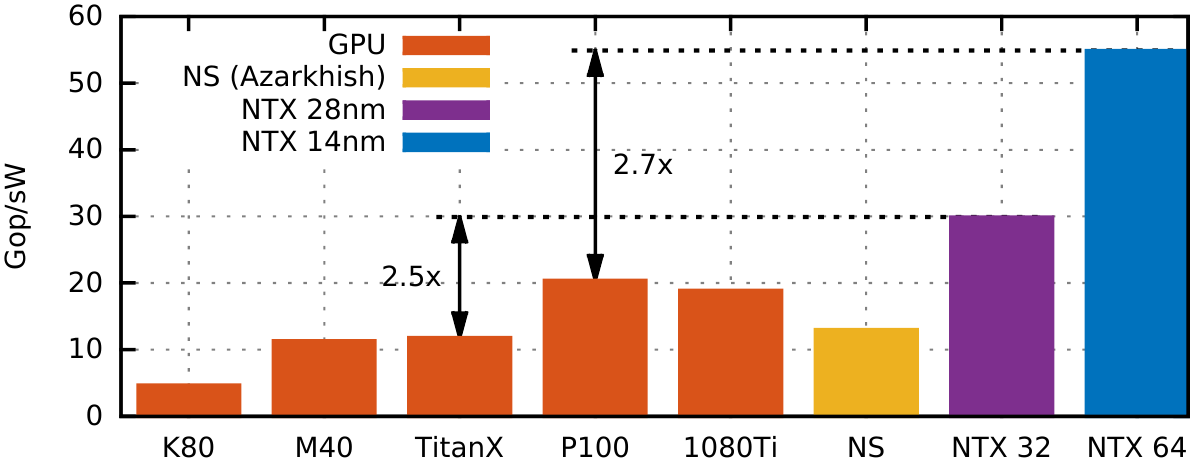}
  \caption{Comparison of energy efficiency when training the networks listed in \tabref{tbl:results:syscmp} (geometric mean), with GPUs, NS \cite{azarkhish2017neurostream}, and the largest NTX configurations that do not require additional \glspl{lim}. NTX\,32 in 28\,nm achieves a $2.5\times$ increase, and NTX\,64 in 14\,nm a $2.7\times$ increase in efficiency over GPUs in similar technology nodes.}
  \label{fig:results:eff_vs_net}
  \vspace{-3mm}
\end{figure}

To compare against other accelerators, we use one training step of AlexNet \cite{Krizhevsky2012}, GoogLeNet \cite{Szegedy2015}, Inception v3 \cite{Szegedy2016}, three variants of ResNet \cite{he2016deep}, and a \gls{lstm} with 512 inputs and hidden states as workload. \tabref{tbl:results:syscmp} and \figref{fig:results:eff_vs_net} provides an overview of the compared architectures.

\begin{revised}
To our knowledge there are two other custom accelerators besides NeuroStream \cite{azarkhish2017neurostream} that claim support for training at precisions similar to ours: DaDianNao \cite{luo2017dadiannao} and ScaleDeep \cite{venkataramani2017scaledeep}. Both provide much less memory relative to their computational power than GPUs, NeuroStream, and NTX. To compare on a system level, we estimate the efficiency of these accelerator including additional DRAM to hold training data, as described in \secref{sec:relwork:train}. In this case, DaDianNao has an efficiency of \SI{65.8}{\GOPsW} with fixed-point arithmetic, which is identical to the computationally equivalent NTX~128. ScaleDeep has an efficiency of \SI{100.8}{\GFLOPsW} which is $1.3\times$ higher than NTX~512, the largest configuration considered by us.
\end{revised}



GPUs are currently the accelerator of choice to train \glspl{dnn}. Our architecture can achieve significantly higher energy efficiency than a GPU at a comparable technology node (see \figref{fig:results:eff_vs_net}). Considering the largest NTX configurations that do not require additional \glspl{lim}, we achieve an efficiency increase of $2.5\times$ from \SI{11.8}{\GOPsW} to \SI{29.9}{\GOPsW} in \SI{28}{\nano\meter}, and an increase of $2.7\times$ from \SI{20.4}{\GOPsW} to \SI{54.9}{\GOPsW} in \SI{14}{\nano\meter}. \revb{Compared to the GPU power analysis and model published in \cite{hong2010integrated}, NTX spends a larger fraction of power in the FPUs, namely 14\% versus 4.8\%. Assuming an FMA requires the same energy per item in similar technology nodes, this increase corresponds to the observed efficiency increase and gives an intuition of why NTX outperforms GPUs. This is in part due to the absence of caches in NTX and the GPU's significant idle power. See \figref{fig:results:power_breakdown}.}

\subsection{Deployed Silicon}
\label{sec:results:deployed_silicon}

\begin{figure}
  \centering
  \includegraphics[width=\linewidth]{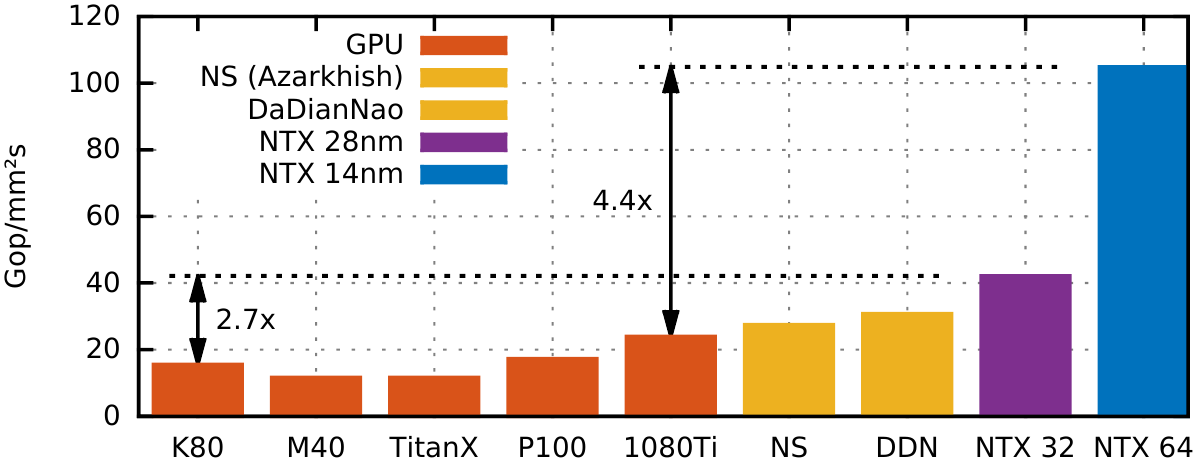}
  \revb{\caption{Comparison of the \si{\GOPs} of compute performance per deployed area of silicon, for GPUs, NS \cite{azarkhish2017neurostream}, and the largest NTX configurations that do not require additional \glspl{lim}. NTX\,32 in 28\,nm achieves a $2.7\times$ increase, and NTX\,64 in 14\,nm a $4.4\times$ increase in area efficiency over GPUs in similar technology nodes.}}
  \label{fig:results:deployed_si}
\end{figure}

One unique key benefit of our architecture is that it leverages existing unused silicon area. This incurs almost no additional costs, since we assume the \glspl{hmc} to be already present in the system as main memory of the CPU, and manufacturing costs of the spare silicon area is the same regardless of whether it is being used. This allows us to deploy up to 32 processing clusters in \SI{28}{\nano\meter} and 64 processing clusters in \SI{14}{\nano\meter} with no additional silicon needed.

\figref{fig:results:deployed_si} compares the \si{\GOPs} of compute performance per deployed amount of silicon for NTX and GPUs. Our solution requires $4.4\times$ less area to achieve the same compute performance as a GPU. Even more when one considers that the chosen 32 and 64 cluster configurations can fit into the aforementioned unused silicon, their cost is virtually zero. This sets our solution apart from ScaleDeep, DaDianNao, and other GPUs, which require significant silicon overhead.

\begin{revised}
\subsection{Scaling to multiple HMCs}
\label{sec:results:multiple_hmcs}

\begin{figure}
  \centering
  \makebox(0,0)[lb]{\textbf{(a)}}\includegraphics[width=\dimexpr0.45\linewidth-3pt]{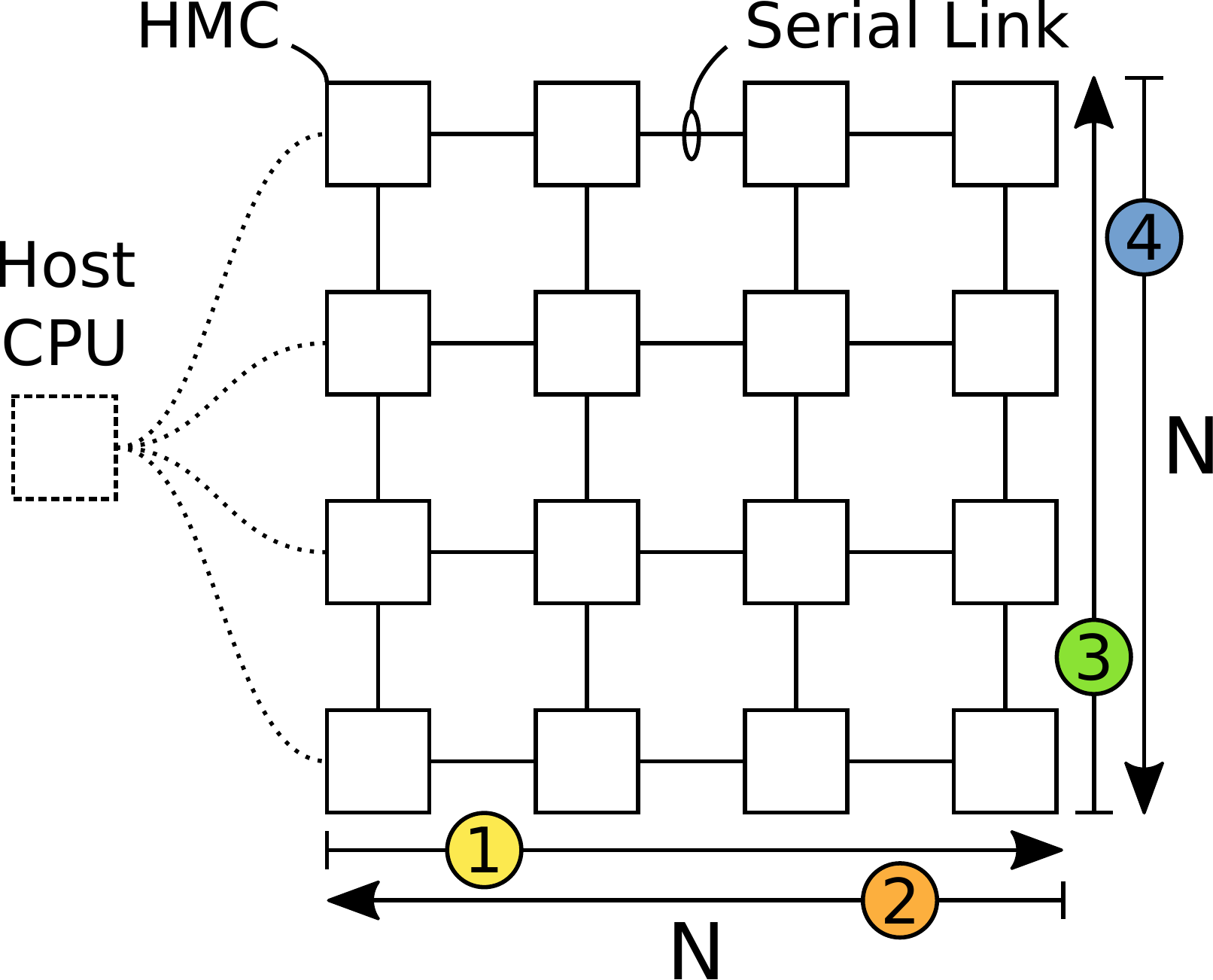}\hfill%
  \makebox(0,0)[lb]{\textbf{(b)}}\includegraphics[width=\dimexpr0.55\linewidth-3pt]{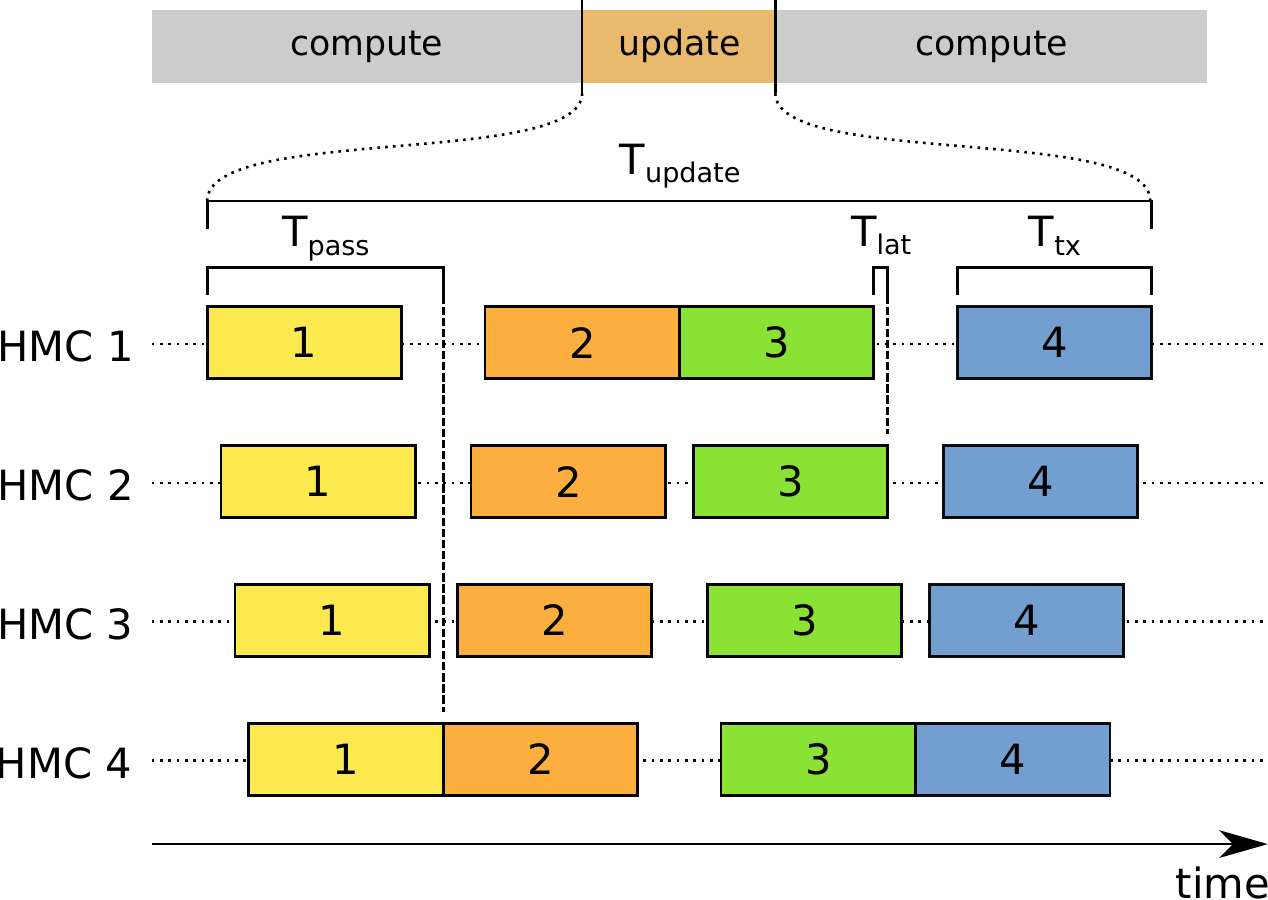}\par
  \medskip
  \makebox(0,0)[lb]{\textbf{(c)}}\includegraphics[width=\linewidth]{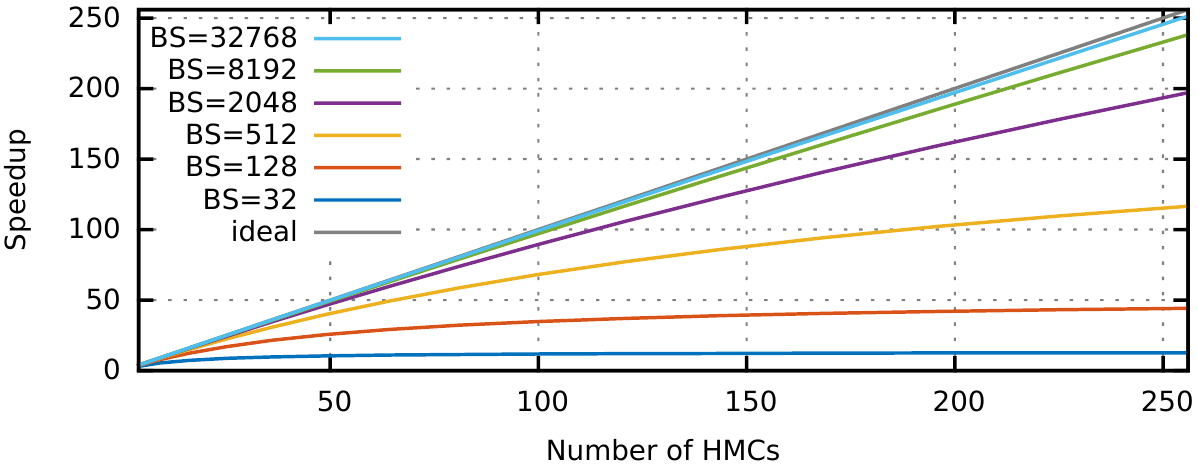}\par
  \medskip
  \makebox(0,0)[lb]{\textbf{(d)}}\includegraphics[width=\linewidth]{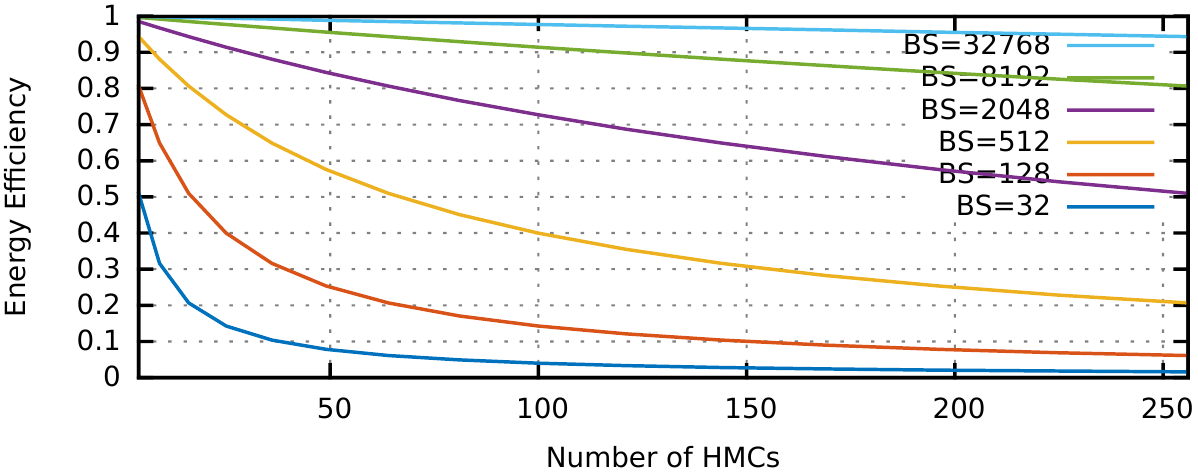}\par
  \caption{\rev{Scaling behavior of data-parallel training on a mesh of \glspl{hmc}. (a) depicts the square mesh with $N^2$ cubes. The arrows indicate the four phases and wave direction of the global weight update. (b) shows the time line corresponding to the mesh for the compute and, in more detail, update phases. (c) outlines the speedup for different mesh and total batch sizes. (d) shows the corresponding energy efficiencies. Larger batch sizes help amortize the cost of the global weight update.}}
  \label{fig:hmc_mesh}
\end{figure}





We investigate the scaling behavior of NTX~64 and organize the \glspl{hmc} in a square mesh of different side lengths $N$ as depicted in \figref{fig:hmc_mesh}a. Each link operates at \SI{60}{\giga\byte\per\second} \cite{HmcV21}. We leverage data parallel training to distribute computation across the \glspl{hmc} in the mesh, which is also commonly done on GPUs \cite{wu2015deep}. Each \gls{hmc} computes its local weight update first. The global update is then performed in four waves as a horizontal followed by a vertical systolic average which can be performed in a streaming fashion.

We assume the weight update to be \SI{300}{\mega\byte}, which takes $T_\text{tx} = \SI{4.88}{\milli\second}$ to transmit. Each cube takes \SI{104}{\micro\second} to compute the average, which is negligible compared to $T_\text{tx}$. Furthermore the internal bandwidth of the \gls{hmc} is much larger than the \SI{120}{\giga\byte} required by the two serial links active in parallel during streaming operation. We assume a latency of $T_\text{lat} = \SI{20}{\micro\second}$ inside the cube, which is a very conservative estimate. The time taken for one of the four waves described above is then
\begin{align}
T_\text{pass} &= T_\text{tx} + N\cdot T_\text{lat}
\end{align}
Since $T_\text{lat}$ is small relative to $T_\text{tx}$, the number of cubes in the mesh has only little influence on this time. For a very large mesh of $N=16$ (256 \glspl{hmc}) $T_\text{pass} = \SI{5.20}{\milli\second}$. Since four such passes are necessary, the total time to perform the weight update across the mesh is
\begin{align}
T_\text{update} &= 4\cdot T_\text{pass} = \SI{20.8}{\milli\second}
\end{align}
A time diagram of such an update is depicted in \figref{fig:hmc_mesh}b. The time required by the mesh to calculate the local weight update is
\begin{align}
T_\text{step} &= \SI{8.69}{\milli\second} \cdot \ifrac{L_B}{N^2}
\end{align}
where $L_B$ is the total batch size across all cubes. This yields a total execution time for one batch across the mesh of $T_\text{total} = T_\text{update} + T_\text{step}$. A single \gls{hmc} would perform the same computation in $T_\text{single} = \SI{8.69}{\milli\second} \cdot L_B$. \figref{fig:hmc_mesh}c depicts the speedup. At a batch size of 8192, a system with 64 \glspl{hmc} achieves almost perfect speedup of $62.8\times$ (98\% parallel efficiency), and 144 \glspl{hmc} achieve $138\times$ (95.8\% parallel efficiency).

Regarding energy efficiency we consider two operating modes of the cubes: During the global mesh update the serial links and clusters are active. We assume the four serial links to consume $P_\text{link} = \SI{8}{\watt}$ \cite{azarkhish2017neurostream}. The energies to compute a wave pass and to power-cycle the serial links \cite{HmcV21} are
\begin{align}
E_\text{pass} &= T_\text{pass} \cdot ( \SI{21}{\watt} + P_\text{link} ) = \SI{150.9}{\milli\joule} \\
E_\text{pwrud} &= 2\cdot P_\text{link} \cdot \SI{50}{\milli\second} = \SI{800}{\milli\joule}
\end{align}
The energies spent per \gls{hmc} for the global and local weight updates are
\begin{align}
E_\text{update} &= 4\cdot E_\text{pass} + E_\text{pwrud} = \SI{1.403}{\joule}\\
E_\text{step} &= T_\text{step} \cdot \SI{21}{\watt} \cdot N^2
\end{align}
and the overall energy for one batch across the mesh requires
\begin{align}
E_\text{total} = ( E_\text{update} + E_\text{step} ) \cdot N^2
\end{align}
A single \gls{hmc} would require $E_\text{single} = T_\text{single} \cdot \SI{21}{\watt}$ for the same task. At a batch size of 8192 as above, 64 \glspl{hmc} achieve an energy efficiency of 94.3\%, 144 \glspl{hmc} achieve 88.1\%.
\end{revised}

\subsection{Savings at Data Center Scale}
\label{sec:results:data_center}

Computing at a data center scale incurs a significant energy and cost overhead over the raw hardware's power consumption. This is among other factors due to the required air conditioning and cooling. A standard measure for this overhead is the \gls{pue} \cite{iso30134}, the ratio of the power consumed by a data center to the power consumed solely by its compute units:

$$\eta_\text{pue} = \frac{P_\text{total}}{P_\text{compute}}$$

\rev{Data centers are reported to have $\eta_\text{pue}=1.12$ \cite{gao2014machine}.} The figure depends heavily on the local climate and usually only the winter months' numbers are published. We assume an average $\eta_\text{pue}=1.2$. We consider a NVIDIA DGX-1 server with two Intel Xeon CPUs and eight Tesla P100 cards. One such unit consumes \SI{3.2}{\kilo\watt} of power, \SI{2.4}{\kilo\watt} of which are due to the GPUs. We assume DDR4 DRAM to consume \SI{6}{\watt} per \SI{16}{\giga\byte} of storage under full load \cite{angelli2014ddr4power}. We investigate two different approaches of replacing the GPUs of the system with NTX-augmented \glspl{hmc}. \rev{Consider that the \SI{512}{\giga\byte} of system memory requires 256 chips distributed across the DIMM modules if built from \SI{16}{\giga\bit} DRAM chips. An 8 GB HMC is roughly equivalent to 4 such chips, so the same system built from \glspl{hmc} would comprise 64 memory cubes.} 

\subsubsection{Same Peak Compute}

\begin{figure}
  \centering
  \includegraphics[width=\linewidth]{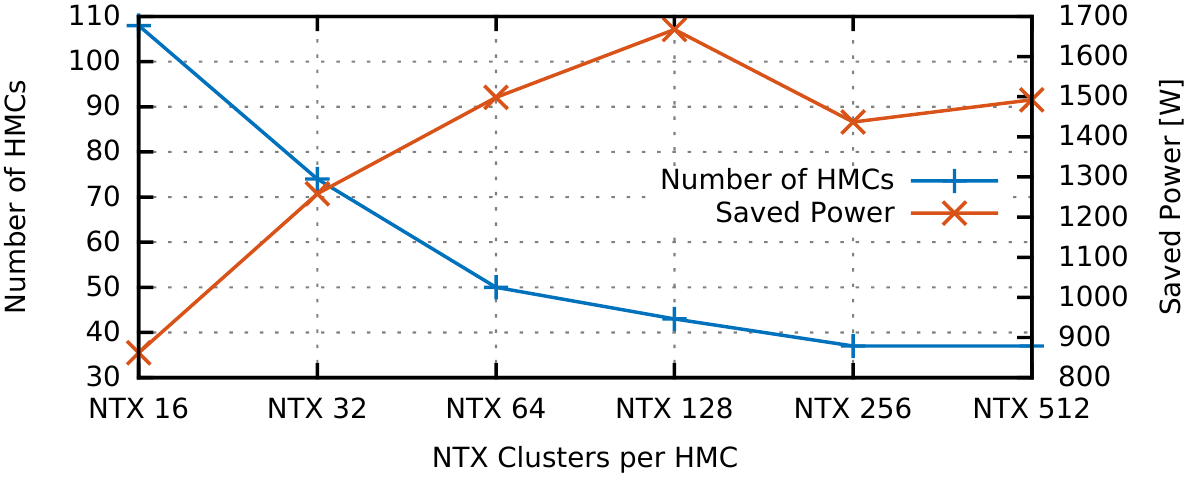}
  \caption{Number of \glspl{hmc} required to meet a compute of \SI{84.8}{\TFLOPs} for different numbers of NTX clusters per \gls{hmc}, and the corresponding power savings over 8 GPUs achieving the same compute capability.}
  \label{fig:datacenter_same_compute}
\end{figure}

\begin{figure}
  \centering
  \includegraphics[width=\linewidth]{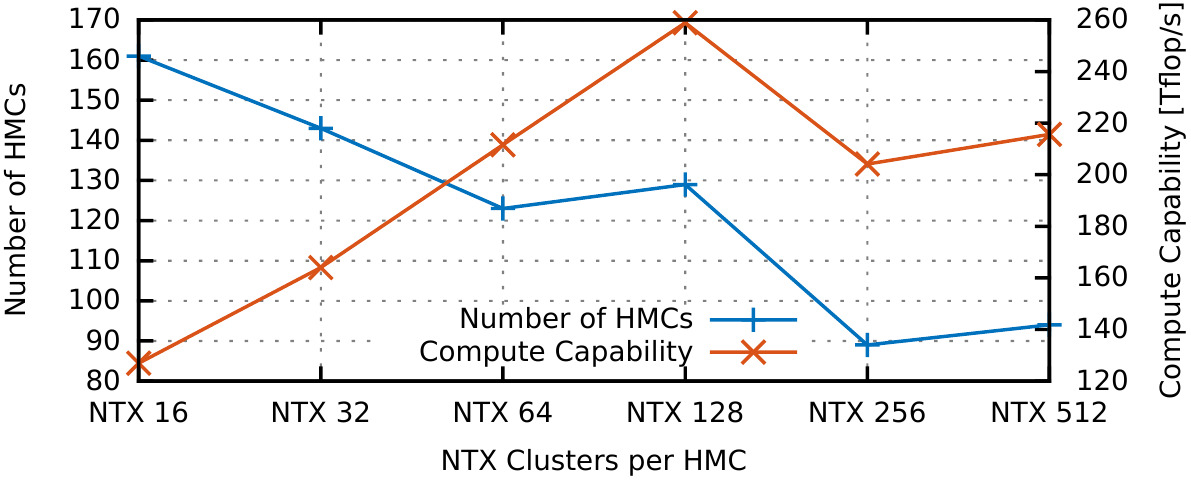}
  \caption{Number of \glspl{hmc} that can be deployed with a power budget of \SI{2.4}{\kilo\watt} for different numbers of NTX clusters per \gls{hmc}, and the corresponding compute capability.}
  \label{fig:datacenter_same_tdp}
\end{figure}

The P100 cards achieve a combined peak compute of \SI{84.8}{\TFLOPs}. \figref{fig:datacenter_same_compute} shows the number of \glspl{hmc} required to match this performance, and the achievable energy savings, with different NTX configurations per cube. The 43 \glspl{hmc} with NTX~128 required to achieve the same compute power consume only \SI{860}{\watt}, saving \SI{2.4}{\kilo\watt} of GPU power and an additional \SI{128}{\watt} of DRAM power, for an overall reduction of $2.1\times$. With a \gls{pue} of 1.2 this translates to \SI{1868}{\kilo\watt} of saved power, which at an energy price of \SI{0.1104}{\$\per\kilo\watt\hour} \cite{eia2017energyprice} is \SI{1808}{\$} per year and server.

\subsubsection{Same Thermal Design Power}

\figref{fig:datacenter_same_tdp} shows the number of \glspl{hmc} that can be deployed within the \SI{2.4}{\kilo\watt} GPU power budget of the DGX-1. 129 \glspl{hmc} with NTX~128 are capable of achieving \SI{258.9}{\TFLOPs} in total, a $3.1\times$ improvement over the P100.

%% file: tables/network_memory.tex
\begin{table}
\begin{threeparttable}
\caption{\rev{Memory footprint of the parameters and intermediate activations of select \glspl{dnn}. The two last columns show the total memory requirement for training with batch size 1 and higher. [\si{\mega\byte}]}}
\label{tbl:network_memory}
\begin{tabularx}{\linewidth}{@{}X k{3.1} k{3.1} k{3.1} k{3.1} @{}}
\toprule
    \textbf{Network} &
    {\textbf{Param.}} &
    {\textbf{Interm. Act.}} &
    {\textbf{BS=1}\tnote{$\dag$}} &
    {\textbf{BS>1}\tnote{$\dag$}} \\
\midrule
    AlexNet      \cite{Krizhevsky2012} & 232.5 &   6.0 & 238.5 & 471.0 \\
    GoogLeNet    \cite{Szegedy2015}    &  26.7 &  46.5 &  73.2 &  99.8 \\
    Inception v3 \cite{Szegedy2016}    &  90.8 &  99.2 & 190.0 & 280.8 \\
    ResNet-34    \cite{he2016deep}     & 176.2 &  28.3 & 204.5 & 380.6 \\
    ResNet-50    \cite{he2016deep}     & 174.6 &  67.1 & 241.7 & 416.3 \\
    ResNet-152   \cite{he2016deep}     & 306.4 & 154.4 & 460.7 & 767.1 \\
\bottomrule
\end{tabularx}
\begin{tablenotes}
    \item[\dag] Batch Size
\end{tablenotes}
\end{threeparttable}
\vspace{-3mm}
\end{table}

%% file: tables/nst_comp.tex
\begin{table}[!t]
\newcommand\resindent{\hspace{0.2em}--\hspace{0.5em}}
\begin{threeparttable}
\caption{Architecture comparison in 28\,nm FD-SOI between NTX (this work) and the inference architecture NS \cite{azarkhish2017neurostream}.}
\label{tbl:nst_comp}
\begin{tabularx}{\linewidth}{@{}X k{3.3} k{3.3} k{3.3} @{}}
\toprule
                               & {\textbf{NS}}          & {\textbf{NTX}}  & {\textbf{NTX}} \\
    {\textbf{Figure of Merit}} & {\cite{azarkhish2017neurostream}} & {``small''}     & {``big''} \\
\midrule
    Clusters/Cores/Accelerators & {16/4/2} & {16/1/8} & {64/1/8} \\
    Cluster/Accelerator Freq.\@ [\si{\GHz}]
        & {1.0/1.0}
        & {0.75/1.5}
        & {0.75/1.5} \\
    Peak Performance [\si{\giga\op\per\s}]
        & {256}
        & {384}
        & {1536} \\
    Core Efficiency [\si{\GOPsW}]
        & {116}
        & {97}
        & {97} \\

\midrule
    \textbf{Area} [\si{\milli\metre\squared}] & \\
    Clusters Logic
        & 4.48
        & 5.38
        & 21.5 \\
    Clusters Memory
        & 4.8
        & 5.1
        & 19.5 \\
    Total
        & 9.3
        & 10.5
        & 41.0 \\

\midrule
    \textbf{Power} [\si{\watt}] & \\
    Clusters Logic
        & 1.10
        & 2.31
        & 9.24 \\
    Clusters Memory
        & 1.10
        & 1.65
        & 6.60 \\
    HMC without Clusters
        & 9.00
        & 9.14
        & 12.9 \\
    Total
        & 11.2
        & 13.1
        & 28.7 \\

\midrule
    \textbf{Inference (GoogLeNet, 1 image)} \\
    Execution Time [\si{\milli\s}]
        & 14.0
        & 11.3
        & 2.83 \\
    Avg./Peak Bandwidth {[\si{\giga\byte\per\s}]}
        & {14.4/51.2}
        & {17.8/57.6}
        & {71.0/230}  \\
    Efficiency {[\si{\GOPsW}]}
        & \ubold 20.3
        & \ubold 21.4
        & \ubold 39.1 \\

\midrule
    \textbf{Training (GoogLeNet, 1 image)} \\
    Execution Time [\si{\milli\s}]
        & 56.8
        & 34.8
        & 8.69 \\
    Avg./Peak Bandwidth {[\si{\giga\byte\per\s}]}
        & {11.3/51.2}
        & {18.5/57.6}
        & {74.0/231}  \\
    Efficiency {[\si{\GOPsW}]}
        & \ubold 15.0
        & \ubold 21.0
        & \ubold 38.3 \\

\bottomrule
\end{tabularx}
\end{threeparttable}
\end{table}

%% file: tables/system_comp.tex
\begin{table*}
\newcommand\nbold{}
\begin{threeparttable}
\caption{Comparison between different configurations of the architecture proposed in this work, related custom accelerators, and GPUs. The energy efficiencies reported are with respect to training different \glspl{dnn} and an LSTM.}
\label{tbl:results:syscmp}
\begin{tabularx}{\linewidth}{@{}X llk{3.1}ck{1.2}k{3.3}c *{8}{k{3.1}}@{}}
\toprule

  \textbf{Platform} &
  \multicolumn{7}{c}{\textbf{Characteristics}} &
  \multicolumn{8}{c@{}}{\textbf{Energy Efficiency} [\si{\GOPsW}]} \\

  \cmidrule(lr){2-8}
  \cmidrule(l){9-16}

  &
  \rotatebox{90}{Logic [\si{\nano\metre}]} &
  \rotatebox{90}{DRAM [\si{\nano\metre}]} &
  \rotatebox{90}{Area [\si{\milli\meter\squared}]} &
  \rotatebox{90}{\gls{lim}} &
  \rotatebox{90}{Freq. [\si{\giga\hertz}]} &
  \rotatebox{90}{Peak \si{\tera\op\per\second}} &
  \rotatebox{90}{Arithmetic \tnote{\dag\dag}} &
  \rotatebox{90}{AlexNet \cite{Krizhevsky2012}} &
  \rotatebox{90}{GoogLeNet \cite{Szegedy2015}} &
  \rotatebox{90}{Incep. v3 \cite{Szegedy2016}} &
  \rotatebox{90}{ResNet 34 \cite{he2016deep}} &
  \rotatebox{90}{ResNet 50 \cite{he2016deep}} &
  \rotatebox{90}{ResNet 152 \cite{he2016deep}} &
  \rotatebox{90}{\textbf{Geom. Mean}} &
  \rotatebox{90}{LSTM\tnote{$\mathsection$}} \\

\midrule

  \multicolumn{7}{@{}l}{\textbf{This Work}} \\
  NTX (16$\times$)  & 28 & 50 & \ubold  10.5 & 0 & 2.30 & 0.589 & (a) &  19.7 &  23.6 &  24.1 &  21.6 &  21.3 &  23.5 & \ubold  22.3 &  29.9 \\
  NTX (32$\times$)  & 28 & 50 & \ubold  20.7 & 0 & 1.70 & 0.870 & (a) &  26.3 &  31.6 &  32.3 &  28.9 &  28.5 &  31.4 & \ubold  29.9 &  40.0 \\
  NTX (64$\times$)  & 28 & 50 & \ubold  41.0 & 1 & 1.30 & 1.331 & (a) &  34.0 &  40.8 &  41.7 &  37.3 &  36.8 &  40.6 & \ubold  38.6 &  51.6 \\
  NTX (16$\times$)  & 14 & 30 & \ubold   4.2 & 0 & 3.08 & 0.788 & (a) &  28.8 &  34.6 &  35.4 &  31.6 &  31.2 &  34.4 & \ubold  32.8 &  43.8 \\
  NTX (32$\times$)  & 14 & 30 & \ubold   8.3 & 0 & 2.24 & 1.219 & (a) &  38.0 &  45.6 &  46.7 &  41.7 &  41.2 &  45.4 & \ubold  43.2 &  61.3 \\
  NTX (64$\times$)  & 14 & 30 & \ubold  16.4 & 0 & 1.68 & 1.720 & (a) &  48.3 &  58.0 &  59.3 &  53.0 &  52.3 &  57.7 & \ubold  54.9 &  73.1 \\
  NTX (128$\times$) & 14 & 30 & \ubold  32.8 & 1 & 0.98 & 2.007 & (a) &  57.9 &  69.5 &  71.0 &  63.4 &  62.6 &  69.1 & \ubold  65.8 &  90.1 \\
  NTX (256$\times$) & 14 & 30 & \ubold  65.6 & 2 & 0.56 & 2.294 & (a) &  65.5 &  78.6 &  80.4 &  71.8 &  70.9 &  78.2 & \ubold  74.4 & 111.2 \\
  NTX (512$\times$) & 14 & 30 & \ubold 131.2 & 3 & 0.28 & 2.294 & (a) &  69.1 &  82.9 &  84.8 &  75.7 &  74.8 &  82.5 & \ubold  78.5 & 116.9 \\

\midrule

  \multicolumn{7}{@{}l}{\textbf{Custom Accelerators}} \\
  NS (16$\times$) \cite{azarkhish2017neurostream} & 28 & 50    & \ubold  9.3 & {---} & 1.0 & 0.256 & (a) & 10.2            & 15.1            & 14.6  & 13.1             & 12.9  & 14.2  & \ubold 13.0             & {---} \\
  DaDianNao \cite{luo2017dadiannao}               & 28 & 28    & \ubold 67.7 & {---} & 0.6 & 2.09  & (b) & {---}           & {---}           & {---} & {---}            & {---} & {---} & \ubold 65.8{\tnote{*}}  & {---} \\
  ScaleDeep \cite{venkataramani2017scaledeep}     & 14 & {---} & {---}       & {---} & 0.6 & 680   & (c) & 87.7{\tnote{*}} & 83.0{\tnote{*}} & {---} & 139.2{\tnote{*}} & {---} & {---} & \ubold 100.8{\tnote{*}} & {---} \\

\midrule

  \multicolumn{7}{@{}l}{\textbf{GPUs}} \\
  Tesla K80 \tnote{\dag}    & 28 & 40\tnote{$\times$} & \ubold 561 & {---} & 0.59 & 8.74 & (a) & {---} & 4.5  & 3.5   & {---} & 3.7   & 8.8   & \ubold  4.7 & {---} \\
  Tesla M40 \tnote{\dag}    & 28 & 30\tnote{$\times$} & \ubold 601 & {---} & 1.11 & 7.00 & (a) & {---} & 11.3 & {---} & {---} & {---} & {---} & \ubold 11.3 & 15.6  \\
  Titan X \tnote{\ddag}     & 28 & 30\tnote{$\times$} & \ubold 601 & {---} & 1.08 & 7.00 & (a) & 12.8  & 9.9  & {---} & 17.6  & 8.5   & 12.2  & \ubold 11.8 & {---} \\
  Tesla P100 \tnote{\dag}   & 16 & 21\tnote{$\circ$}  & \ubold 610 & {---} & 1.3  & 10.6 & (a) & {---} & 19.8 & 19.5  & {---} & 18.6  & 24.18 & \ubold 20.4 & {---} \\
  GTX 1080 Ti \tnote{\ddag} & 16 & 20\tnote{$\times$} & \ubold 471 & {---} & 1.58 & 11.3 & (a) & 20.1  & 16.6 & {---} & 27.6  & 13.4  & 19.56 & \ubold 18.9 & {---} \\

\bottomrule
\end{tabularx}
\begin{tablenotes}
    \item[\dag] Inception/ResNet: batch size 64 with TensorFlow/cuDNN 5.1 \cite{TfBench2017}; GoogLeNet: batch size 128 with Torch/cuDNN 5.1 \cite{murphy2017}
    \item[\ddag] All nets: batch size 16 Torch/cuDNN 5.1 \cite{johnson2017}
    \item[$\mathsection$] 512 inputs and hidden states, batch size 32 for NTX and 64 for GPU \cite{appleyard2016optimizing}
    \item[$\times$] GDDR5 and GDDRX5, process node estimated based on GPU release year
    \item[$\circ$] HBM2
    \item[*] \rev{Estimated system efficiency including DRAM, see \secref{sec:relwork:train}}
    \item[\dag\dag] (a) floating-point 32\,bit, (b) fixed-point 16/32\,bit, (c) floating-point 16/32\,bit
\end{tablenotes}
\end{threeparttable}
\end{table*}

%% file: sec_relwork.tex
\section{Related Work}
\label{sec:relwork}


Acceleration of \glspl{dnn}, in particular the forward pass, is a well researched field with a rich literature. Goodfellow, et al.\@ \cite{Goodfellow2016} provide a good coverage of the mathematical background of Deep Learning. An overview of techniques for efficient DNN inference and the involved challenges can be found in \rev{\cite{Sze2017}}.

\subsection{Accelerators for Inference}
\label{sec:relwork:infer}

\begin{revised}
Architectures to accelerate the inference process of \gls{cnn} have been studied extensively in literature. FPGA-based accelerators report energy efficiencies on the order of \SI{10}{\GOPsW} and usually rely on fixed-point arithmetic and less than 32\,bit precision \cite{Gokhale2017}. ASIC-based accelerators porivde efficiencies on the order of \SI{1000}{\GOPsW} at reduced precisions, for example Google's TPU \cite{Jouppi2017} which uses 8\,bit arithmetic, or \cite{Andri2017} with 1\,bit. Near-memory inference architectures embedded in the logic die of an \gls{hmc} have also been investigated, for example the 2D accelerator array presented in \cite{Gao2017} which uses 16\,bit fixed-point arithmetic and achieves up to \SI{450}{\GOPsW}, or the clustered many-core architecture \cite{azarkhish2017neurostream} which is based on 32\,bit \gls{fp} co-processors and achieves up to \SI{22.5}{\GFLOPsW}. Certain architectures such as \cite{chen2014diannao} employ a distributed memory model, where the entire network's parameter are stored on chip. This becomes increasingly difficult with modern networks that require hundreds of \si{\mega\byte} \cite{Krizhevsky2012, Szegedy2015, Szegedy2016, he2016deep}, and the network to be trained is tightly bound to the number of chips that can be interconnected. We furthermore observe that due to the vast difference in energy spent for computation and data transfer, it is only meaningful to compare architectures that use the same arithmetic precision and bit width.
\end{revised}

\subsection{Accelerators for Training}
\label{sec:relwork:train}

\begin{revised}
We observe that much fewer architectures have been proposed to cover the training aspect of \glspl{dnn}. Many of the aforementioned architectures are not suitable for this since they lack the ability or memory capacity to store intermediate activations, e.g. due to optimizations in the data path, or the precision and dynamic range for the training to converge.

The NeuroCube \cite{kim2016neurocube} is based on 16\,bit fixed-point MAC units which are capable of performing the necessary computations, but it is unclear if training of modern deep networks converges at this precision and dynamic range. NTX surpasses NeuroCube's efficiency of \SI{7.63}{\GOPsW} because we focus on maximizing the energy spent in the FPU, e.g. by doubling its clock frequency. We furthermore use \gls{vfs} to increase efficiency.

DaDianNao \cite{luo2017dadiannao} uses 64 chips to perform training at 32\,bit fixed-point precision with \SI{2.3}{\giga\byte} of distributed memory. A single chip with \SI{2.1}{\TOPs} and \SI{36}{\mega\byte} is roughly equivalent to NTX~128 with \SI{2}{\TFLOPs} and \SI{16}{\mega\byte} but without the HMC. In this setting NTX achieves a $1.9\times$ better core efficiency of \SI{250.9}{\GFLOPsW} despite its \gls{fp} arithmetic. Considering an estimated \SI{15.8}{\giga\byte} of DRAM that is needed to match the memory-to-compute density of the DGX-1 puts DaDianNao's system energy efficiency at \SI{65.8}{\GFLOPsW}, assuming that the overall cost of the memory (DRAM modules, memory controller, interconnects) is comparable to that of an \SI{8}{\giga\byte} \gls{hmc} (\SI{1}{\watt\per\giga\byte}).

ScaleDeep \cite{venkataramani2017scaledeep} supports training in 32\,bit \gls{fp} precision at 14\,nm. We estimate its total die area to be \SI{2800}{\square\milli\meter} based on its TDP and the power density of a GPU, which yields \SI{243}{\giga\FLOP\per\milli\meter\squared\second}, around $2.3\times$ more than NTX~64. An entire node achieves \SI{680}{\TFLOPs} but has only \SI{1.17}{\giga\byte} of distributed memory. As such it excludes any form of system DRAM and puts its core efficiency of \SI{420.9}{\GFLOPsW} on par with the \SI{417.0}{\GFLOPsW} achieved by NTX~512. The estimated \SI{5.13}{\tera\byte} of DRAM that is needed to match the memory-to-compute density of the DGX-1 puts ScaleDeep's system power consumption at \SI{6.75}{\kilo\watt} under the same assumptions as DaDianNao, with a system energy efficiency of \SI{100.8}{\GFLOPsW}. It is unclear how much additional energy would be consumed by the processing power required to feed these accelerators with data.

\end{revised}

\subsection{GPUs}
\label{sec:relwork:gpu}

\begin{revised}
GPUs can be seen as the main workhorse of Deep Learning and are commonly used for both inference and training due to their flexibility. Recent implementations on the GTX~780 and GTX~Titan (both featuring a Kepler microarchitecture) reach \SI{1650}{\GFLOPs} at \SI{250}{\watt} and \SI{999}{\GFLOPs} at \SI{240}{\watt}, which corresponds to 6.6 and \SI{4.2}{\GFLOPsW}, respectively \cite{Cavigelli2015b, azarkhish2017neurostream}. Embedded GPUs like the Tegra K1 have lower absolute throughput, but reach a similar energy efficiency of around \SI{7}{\GFLOPsW} \cite{Cavigelli2015b}. The Pascal generation of GPUs offer several features beneficial to \glspl{dnn}, such as \gls{hbm} and 16\,bit \gls{fp} support. Compared to previous generations, the P100 achieves a $2\times$ higher peak throughput of \SI{10.6}{\TFLOPs} and a significantly higher energy efficiency around \SI{20}{\GFLOPsW} \cite{johnson2017, TfBench2017}. The recently introduced Volta generation offers \emph{tensor cores}, a new compute element able to perform 4$\times$4 matrix \glspl{fma} in 16\,bit \gls{fp}, with 16 or 32\,bit outputs in one cycle. These cores promise a $5\times$ increase in Deep Learning performance compared to previous GPU generations \cite{nvidia2017volta}. \revb{Furthermore, GPUs have been shown to be amenable to near-memory processing as well \cite{pattnaik2016scheduling}.}
\end{revised}

%% file: sec_conc.tex
\section{Future Work}
\label{sec:futwork}

\begin{revised}
Our architecture is inherently scalable since the \gls{hmc} standard allows for memory cubes to be interconnected via the serial links \cite{HmcV21}. Mesh arrangements of \glspl{hmc} offer many opportunities and different parallelization techniques \cite{mcmahan2016communication} for training \gls{dnn} should be explored. Moving to \gls{hbm} promises further energy efficiency gains and brings new challenges and design constraints in using the bottom memory controller die. Improvements to the DMA engine in the compute clusters would allow for even more efficient offloading and further ease the load on the RISC-V processor core. Applicability of transprecision and compression techniques offer other interesting angles to be investigated for further gains.
\end{revised}

\section{Conclusion}
\label{sec:conc}

We have presented the streaming \gls{fp} co-processor NTX with a decisive focus on training DNNs. Its data path is built around a fast fused accumulator with full \SI{32}{\bit} precision, which gives it a key advantage over architectures that are based on fixed-point arithmetic or lower \gls{fp} precision. The co-processor is capable of generating three independent address streams from five nested hardware loops, allowing it to traverse structures with up to five dimensions in memory independently. A rich set of arithmetic and logic commands allows it to perform the reductions and matrix/vector operations commonly found in the forward pass, but also the threshold, mask, and scatter operations encountered during the backward pass. We combine eight such co-processors with memory, a control processor, and a DMA unit into a cluster. An efficient offloading scheme frees up resources on the control processor to exert fine-grained control over data movement. The data does therefore not need to be put into memory in a specific, pre-tiled pattern, but can be operated on directly in its canonical and dense form. Integrated into the \gls{lob} of an \gls{hmc}, multiple clusters can exploit the high bandwidth and low accesses latency into DRAM in this near-memory setting. Configurations which fit into the unused area on the \gls{lob} incur virtually zero additional manufacturing costs. NTX scales well to large meshes of \glspl{hmc} and can provide the same compute capability at less power, or more compute capability at the same power.